\documentclass[useAMS,usenatbib]{mn2e}
\usepackage{graphicx}
\usepackage{ulem}
\usepackage{epstopdf}

\def\be{\begin{equation}}
\def\ee{\end{equation}}
\def\ba{\begin{eqnarray}}
\def\ea{\end{eqnarray}}

\def\msun{M_\odot}

\def\ltsima{$\; \buildrel < \over \sim \;$}
\def\simlt{\lower.5ex\hbox{\ltsima}}
\def\gtsima{$\; \buildrel > \over \sim \;$}
\def\simgt{\lower.5ex\hbox{\gtsima}}
\def\etal{{et al.\ }}

\title[Clustered supernovae]
{Evolution of clustered supernovae}
\author[E. O. Vasiliev \etal]
       {Evgenii O. Vasiliev$^{1,2}$\thanks{E-mail:eugstar@mail.ru},
        Yuri A. Shchekinov$^{3,4}$, 
        Biman B. Nath$^{4}$\\
$^1$Southern Federal University, Stachki Ave. 194, Rostov-on-Don, 344090 Russia\\
$^2$Special Astrophysical Observatory of Russian Academy of Sciences, Nizhnii Arkhyz, Karachaevo-Cherkesskaya Republic, 369167 Russia \\
$^3$Lebedev Physical Institute of Russian Academy of Sciences, 53 Leninskiy Ave., 119991, Moscow\\
$^4$Raman Research Institute, Sadashiva Nagar, Bangalore 560080, India\\
}
\begin{document}
\date{Accepted 3004 December 15.
      Received 2004 December 14;
      in original form 2004 December 31}
\pagerange{\pageref{firstpage}--\pageref{lastpage}}
\pubyear{3004}
\maketitle

\label{firstpage}

\begin{abstract}
{We study the merging and evolution of isolated supernovae (SNe) remnants in a stellar cluster into a collective superbubble, with the help of 3-D hydrodynamic simulations. 
We particularly focus on the transition stage when the isolated SNe remnants gradually combine to form a superbubble. 
We find that when the SN rate is high ($\nu_{\rm sn}\sim 10^{-9}$ pc$^{-3}$ yr$^{-1}$), the merging phase lasts for $\sim 
10^4$ yr, for $n=1\hbox{--}10$ cm$^{-3}$, and the merging phase lasts for a longer time ($\sim 0.1$ Myr or more) for lower SN rates ($\nu_{\rm sn}\le 10^{-10}$ pc$^{-3}$ yr$^{-1}$). 
During this transition phase, the growing superbubble is filled with dense and cool fragments of shells and most of the energy is radiated away during this merging process. 
After passing through the intermediate phase, the superbubble eventually settles on to a new power-law wind asymptote that is smaller than estimated in a continuous wind model. This results in a significant (more than {\it several times}) underestimation of the mechanical luminosity needed to feed the bubble.
We determine the X-ray and H$\alpha$ surface brightnesses as functions of time for such merging SNe in a stellar cluster and find that
clusters with high SN rate shine predominantly in soft X-rays and H$\alpha$. In particular, a low  value of the volume averaged H$\alpha$ to H$\beta$ ratio and its large spread can be a good indicator of the transition phase of merging SNe.}
\end{abstract}


%----------------------- Section 1 -------------------------------

\section{Introduction}

\noindent

SNe (Supernovae) explosions and their remnants are believed to play a crucial role in shaping the  evolution of galaxies. It became  clear from the seminal paper by \citet{mckee77} that SN remnants may stay isolated during only a limited period in the beginning,  and afterwards they merge with neighbouring remnants to form a percolating network. Depending on the environment, such merged SNe remnants can build giant supershells \citep[as seen, e.g., in Holmberg galaxies ][]{walter99,stewart00,egorov14,egorov17}, and large scale galactic outflows, called galactic winds. Galactic winds in turn act to enrich the Universe with heavy elements. 

Observations of edge-on galaxies have demonstrated that most  spiral galaxies have 5 to 10 kpc scale gaseous haloes, indicating activity of local outflows in the underlying disks. More recently, observational evidences of existence of huge -- up to 150 to 200 kpc -- metal bearing circumgalactic gaseous coronae have also appeared. Such scales suggest much more powerful outflows -- galactic winds -- to feed the coronae. Observations of absorptions in metal lines in the intergalactic medium at high redshifts ($z=2-6$) show the existence of highly enriched -- up to $\sim Z_\odot$ -- gas. Typical intergalactic scales imply that galaxies develop a powerful stellar feedback process in the form of galactic winds that are able to carry metals through over the intergalactic space. 

When the dynamical effects from collective  SNe are studied in detail, a commonly practice is to introduce the concept of a mechanical luminosity $L=\tau_{\rm sn}^{-1}E_{\rm sn}$, where $\tau_{\rm sn}$ is the time between successive SNe, and $E_{\rm sn}$ is the typical mechanical energy released in a supernova. This assumption implicitly requires that explosion energy from different supernovae add up to produce a collective effect. This in turn means that the remnants from the subsequent supernovae overlap {\it before} becoming radiative \citep{nath2013}. Recent calculations have found that when this condition is not fulfilled, the efficiency of SNe energy to convert into the collective action drops by roughly a factor of ~10 \citep{v15,multi-nath}. 

In general, the dynamics of a sequence of concerted SNe can be understood in terms of the characteristic time scales relevant for an isolated SN remnant and their interplay with the time between sequential explosions. These are basically: 1) the time when the Sedov-Taylor (adiabatic) stage gives way to the radiative stage, 2) when radiative losses significantly decrease the remnant energy in one dynamical time, and 3) the time when the remnant velocity falls below the sound speed in ambient medium. The first time scale is marked by the shock velocity dropping roughly below  $c_h\sim 100$ km s$^{-1}$ and the remnant expanding further with decreasing energy as $E\propto R^{-2}$. The porosity of the remnants whose energy has decreased to half of the explosion energy {$E_{\rm sn}$ is estimated to be $P_{1/2}\sim 2\times 10^{10}\nu_{\rm sn}(E_{\rm sn}/10^{51}\, {\rm erg})n_0$ where $\nu_{\rm sn}$ [pc$^{-3}$ yr$^{-1}$] is the supernova rate density \citep{nath2013}. For a typical galactic environment, one has $P_{1/2}\simlt 1$. The subsequent snowplough expansion of a radiative remnant continues until $v_{\rm shell}\sim c_s$, and the corresponding porosity {$P_{v_{\rm shell}\sim c_s}\sim P_{1/2} \, (c_h/c_s) \sim 10-100$} for warm ($T\sim 10^4$ K) and cold ($T\sim 10^2$ K) gas. This shows that the probability for supernovae to occur within a remnant formed by a previously exploded SN is rather large in typical galactic conditions, particularly during radiative stages. Thus an immediate consequence of this estimate is that the dynamical state and morphology of the interstellar medium in galaxies are mostly governed by supernovae explosions as first recognized by \citet{mckee77}  %McKee \& Ostriker 
\citep[see also more recent discussion by][]{v15,ostriker15}. 
 
It is therefore clear that simplified models of giant expanding supershells or galactic winds based on the assumption a single spherical bubble either from a central wind source with mechanical energy {$L\sim\tau_{\rm sn}^{-1}E_{\rm sn}$} \citep{weisz11,egorov14}, or an instant explosion with energy $\sim N_{\rm sn}E_{\rm sn}$ \citep{walter99,HIholes-ott,simpson05,weisz09} may not be realistic. They may produce inconsistencies between the underlying stellar population and estimated energetics of an expanding flow. In order to correctly evaluate the limits of such inconsistencies, here we perform  a 3D hydrodynamical study of the transition from single isolated SNe explosions to merging remnants and the formation of a collective large scale outflow, focusing mainly on observables accompanying the intermediate transitional stage. 

The paper is organized as follows. In Section 2 we describe the details of the model. In Section 3 we consider two simple models: an ensemble of isolated SNe and multiple SNe exploded from the same point. In Section 4 we present the dynamics of clustered SNe and in Section 5 we describe possible observational consequences. Section 6 summarizes the results.

%----------------------- Section 2 -------------------------------

\section{Model description and numerical setup}

\noindent

We carry out 3-D hydrodynamic simulations (Cartesian geometry) of multiple SNe explosions clustered in space. SNe are distributed uniformly and randomly inside a region with a fixed cluster radius $r_c$. Simulations are performed for a set of $r_c$ ranging from 30 to 90~pc for ambient gas density $n=1$~cm$^{-3}$, and up to 120~pc for $n=10$~cm$^{-3}$; {as a fiducial model we assumed the density rate of SNe $\nu_{\rm sn}=10^{-9}(30~{\rm pc}/r_c)^3$~pc$^{-3}$~yr$^{-1}$, {which} corresponds to one SN per $10^4$~yr in the cluster. We also ran models with the rate of one SN in $10^5$ yr.}  We inject the mass and energy of each SN in a region of radius $r_0=2$ pc. We assume that typical mass and energy of the SN are 10$\msun$ and $10^{51}$~erg. The energy is injected in thermal form. The ambient gas density ranges in $1\hbox{--}10$ cm$^{-3}$, while temperature is $10^4$~K; gas  metallicity is kept constant and equal to the solar value within the whole computational domain. 

Recently \cite{multi-nath} considered a similar model for a dynamical structure growing under the action of clustered multiple supernovae. They mostly focused on the study of dynamical features of the growing structure and related thermodynamic variables, such as the overpressure in the remnant, evolution of the fraction of SNe injected energy retained in thermal and kinetic energies, the fraction of gas removed from the cluster, etc. A major difference between their work and ours is connected to the regime of energy injection into the computational domain. In our case we inject thermal energy into a region of radius $r_0=2$ pc homogeneously, and the time lag between the subsequent instantaneous SNe was chosen to lie between $10^4$ to $10^5$ yr. {In contrast,} \cite{multi-nath} injected the energy smoothly over a sphere of radius $r_{\rm SN}=5$ pc. The smoothing procedure is described by the core $\propto\exp(-[t-t_i]^2/\delta t_{\rm inj}^2)\times\exp(-[{\bf x}-{\bf x}_i]^2/r_{\rm SN}^2)$ with $\delta t_{\rm inj}=\delta t_{\rm SN}/10$, where 
\be 
\delta t_{\rm SN}={\tau_{\rm OB}\over N_{\rm OB}},
\ee
$\tau_{\rm OB}=30$ Myr is the lifetime of an OB-association, and $N_{\rm OB}$ is the number of massive stars (supernovae progenitors) in the association. For their fiducial model with $N_{\rm OB}=100$, this means that the energy injection period is $\Delta t\sim\delta t_{\rm inj}\sim 3\times 10^4$ yr, factor of 3 larger than the time lag between the subsequent SNe in our fiducial model. Moreover, the number of SNe exploded within $\Delta t\sim\delta t_{\rm inj}$ in a volume occupied by one remnant is larger than unity. From this point of view the fiducial model of \citet{multi-nath} corresponds to a continuous energy input, while in our fiducial model the injection is discrete. Asymptotic dynamics and kinematics on times $t>1$ Myr in our simulations is similar. 

%It is readily seen that the ejecta avoids artificial overcooling much longer than the free expansion phase discussed in detail in \citet{roy14} and \citet{multi-nath}. Indeed, gas density and temperature in the ejecta at $t=0$ are $n\simeq 10$ cm$^{-3}$ and $T\simeq 5\times 10^8$ K, with characteristic cooling time $t_c\simeq 10^{14}$ s, while the free expansion phase from $r_c=2$ pc to $r_f\simeq 4$ pc lasts $t_f\simeq 3\times 10^{10}$ s.  

Standard simulations are performed with a physical cell size of 1~pc in the computational domain of $300^3$ pc$^3$. Several simulations are run with a box size of $800^3$ pc$^3$, with a cell size of $0.37$ pc to resolve the free expansion phase, 
as shown in Figure A1. The ejecta avoids artificial overcooling much longer than the free expansion phase (as discussed in detail in \citet{roy14} and \citet{multi-nath}), %. Indeed, gas density and temperature in the ejecta at $t=0$ are $n\simeq 10$ cm$^{-3}$ and $T\simeq 5\times 10^8$ K, with characteristic 
as the cooling time $t_c\simeq 10^{14}$ s %, while the free expansion phase from $r_c=2$ pc to $r_f\simeq 4$ pc lasts 
is much longer than the free-expansion time scale $t_f\simeq 3\times 10^{10}$ s.

The code is based on the unsplit total variation diminishing (TVD) approach that provides high-resolution capturing of shocks and prevents unphysical oscillations. We have implemented the Monotonic Upstream-Centered Scheme for Conservation Laws (MUSCL)-Hancock scheme and the Haarten-Lax-van Leer-Contact (HLLC) method \citep[see e.g.][]{toro99} as approximate Riemann solver. This code has successfully passed the whole set of tests proposed in \citep{klingenberg07}. In order to check convergency, we performed runs with half-size cells. 

Simulations are run with radiative cooling processes with a tabulated non-equilibrium cooling curve fitting the calculated one \citep{v11,v13}. The is obtained for a gas cooling isobarically from $10^8$ down to 10~K.  { Our choice of isobaric cooling rate is reasonable because of the fact that the typical sound crossing time (given the typical temperatures) of a resolution element in our simulation is of order $\sim 1000$~yr, much shorter than the relevant time scales in the problem.} 

The non-equilibrium calculation \citep{v11,v13} includes the kinetics of all ionization states of H, He, C, N, O, Ne, Mg, Si, Fe, as well as kinetics of molecular hydrogen at $T<10^4$~K. { We perform non-equilibrium calculations of the hydrogen ionization fraction and emissivities in Balmer lines and implement them into the hydrodynamic step. The emissivities in Balmer lines have been calculated with making use the CLOUDY package as a subroutine \citep{cloudy}. The ionization fraction is utilized for calculating emission measure and then X-ray spectral intensity \citep[e.g.,][]{kaplan}. For this purpose the approximation for the Gaunt factor is taken from \citep{drainebook}.} The heating rate is assumed to be constant, with a value chosen such as to stabilize the radiative cooling of the ambient gas  at $T=10^4$ K. This stabilization vanishes when the gas density and temperature deviate %a narrow vicinity of 
from the equilibrium state %defined as a relative deviation 
by less than {$1\%$}. 

In order to localize the boundaries of isolated SNe remnants or the collective superbubble, we applied two techniques: the first (used as the fiducial) is based on the fact that any motion in the computational domain is  being driven by a SN, i.e. a gas parcel is identified as a part of a bubble when its velocity squared is greater than a given small value. Even though generally this value should be zero in an unperturbed ambient medium, numerical errors introduce perturbations and we set this threshold 1~km~s$^{-1}$. The second is based on the deviation of gas temperature from the temperature of unperturbed ambient gas, i.e. a gas parcel is located inside a bubble if its temperature is not equal to $T=10^4$ K. In practice, the temperature of the ambient gas is kept within a narrow range around $10^4$~K. The last method can result in a loss of a fraction of cooling gas within the bubble, with temperature in this narrow range. Normally this fraction is negligibly small, but in general the two methods may give different bubble volumes, particularly in the higher density case $n=10$ cm$^{-3}$.

%----------------------- Section 3 -------------------------------

\section{Toy models}

Before describing multiple SN explosions in a cluster we consider several simple models. 

\subsection{Isolated  SN remnants}

%%%%%%%%%%%%%%%%%%%%%%%%%%%%%%%%%%%%%%%%%%%%%%%%%%%%%%%%%%%%%%%%%%%%%
\begin{figure}
\center 
\includegraphics[width=8cm]{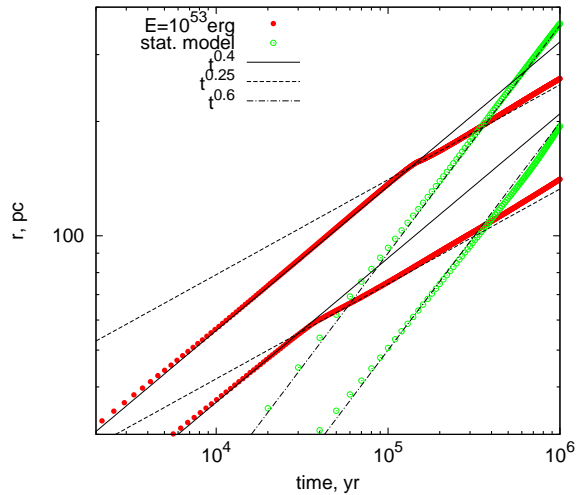}
\caption{
Time dependence of the {\it effective} radius of a cumulative volume of one hundred of isolated SNe exploded with a rate of 1 SN per $10^4$~yr (green { open} symbols), and the bubble radius from a single SN explosion with $E=10^{53}$~erg (red {filled} symbols); {for $E=10^{53}$~erg, the free expansion stage lasts $t_f\sim 10^2$ yr}. The ambient density is 1~cm$^{-3}$ for upper lines and 10~cm$^{-3}$ for lower lines. 
}
\label{fig-1sn100-rad}
\end{figure} 
%%%%%%%%%%%%%%%%%%%%%%%%%%%%%%%%%%%%%%%%%%%%%%%%%%%%%%%%%%%%%%%%%%%%%

For an isolated SN at the very initial stages $t<t_f$, the exponent (that describes the evolution of the bubble radius in the form $R \propto t^\alpha$) turns from the case of free expansion $\alpha\simeq 1$,  through an intermediate `quasi-diffusive' law with $\alpha\simeq 0.45$, (see, Appedix A), to Sedov-Taylor (ST) law $\alpha=2/5$. Fugure~\ref{fig-1sn100-rad} presents the bubble radius from a single SN explosion with $E=10^{53}$~erg (red symbols). In this Figure the expansion has already entered ST phase, and after roughly one radiative time $t_c\sim 10^5$ yr (estimated for a SN with $E=10^{53}$ erg exploded in the medium with $n=1$~cm$^{-3}$) the expansion converges to the momentum driven Oort stage with $\alpha\simeq 1/4$. The intermediate pressure-dominated stage with $\alpha=2/7$ \citep{mckee77,blinn82} continues for a short timescale -- $t_{P}\sim 0.05 t_c$, close to the epoch $t\sim t_c$ (see Appendix B), and remains unresolved on Figure \ref{fig-1sn100-rad}.

Let us consider an ensemble of isolated SNe of different ages. Supernovae explode with a constant rate of 1 SN per $\Delta t$,  each passing through the standard sequence of phases: free expansion, adiabatic (Sedov-Taylor) expansion, and finally the radiative (snowplough) phase. After $k$ explosions the total volume occupied by all remnants is $V(t)=\Sigma_{i=1}^k V_i[t-(i-1)\Delta t]$. For large $k$ the majority of SNe in the ensemble are at the radiative phase and  $V(t)\propto t^{3\alpha+1}$, where the shell radius of a single SN at the radiative stage $r \sim t^\alpha$ with $\alpha = 1/4$ is assumed. It is seen  that the expansion law  asymptotically reaches the wind regime with the {\it effective} radius $R(t)=V^{1/3}\sim t^{0.6}$, nearly coincident with the radius of a wind driven bubble \citep{avedisova,castor}. Throughout the paper, we will denote by {\it effective} radius the size defined as $R=V^{1/3}$. 

Figure~\ref{fig-1sn100-rad} depicts the {\it effective} radius of the ``collective remnant'' of $100$ isolated SNe exploded with the rate of 1 SN per $10^4$~yr (green symbols) and the remnant radius from a single enhanced (with the energy $E=10^{53}$~erg) SN explosion (red symbols) for comparison. Differences between the two cases are clearly seen: while multiple SNe exploding sequentially with a constant rate settle on to the wind mode, a single SN with an enhanced energy passes through the Sedov-Taylor to Oort stage and eventually occupies a lower volume. 

\subsection{Centred multiple SN explosions}

%%%%%%%%%%%%%%%%%%%%%%%%%%%%%%%%%%%%%%%%%%%%%%%%%%%%%%%%%%%%%%%%%%%%%
\begin{figure*}
\center 
\includegraphics[width=8cm]{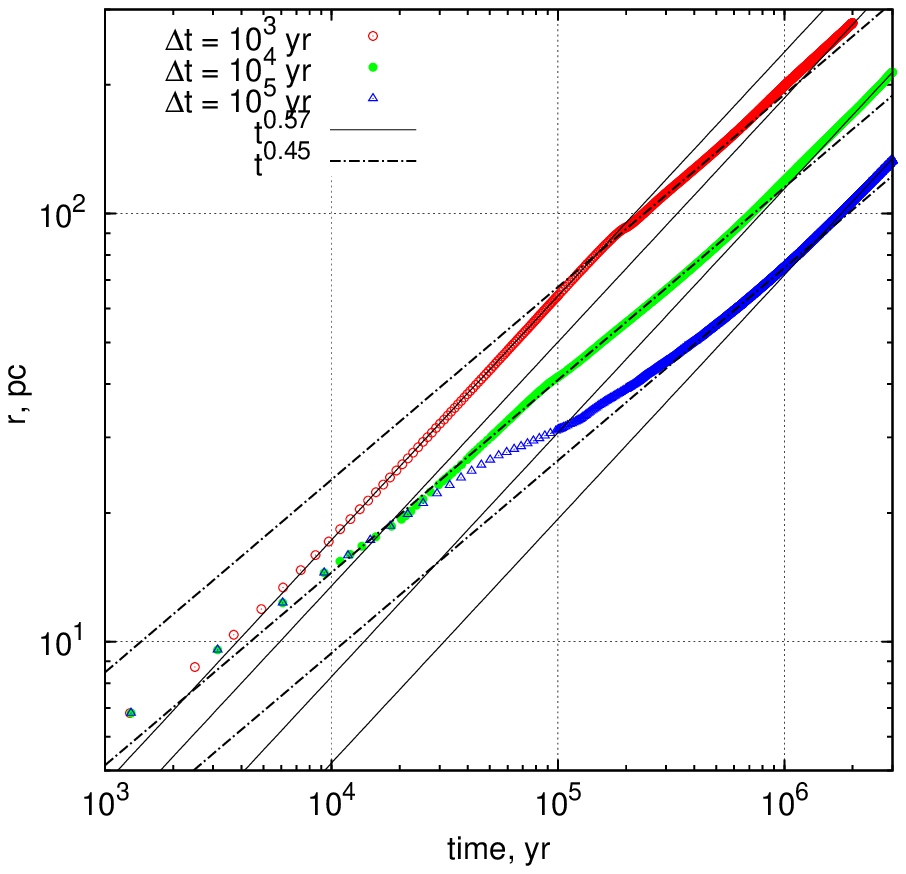}
\includegraphics[width=8cm]{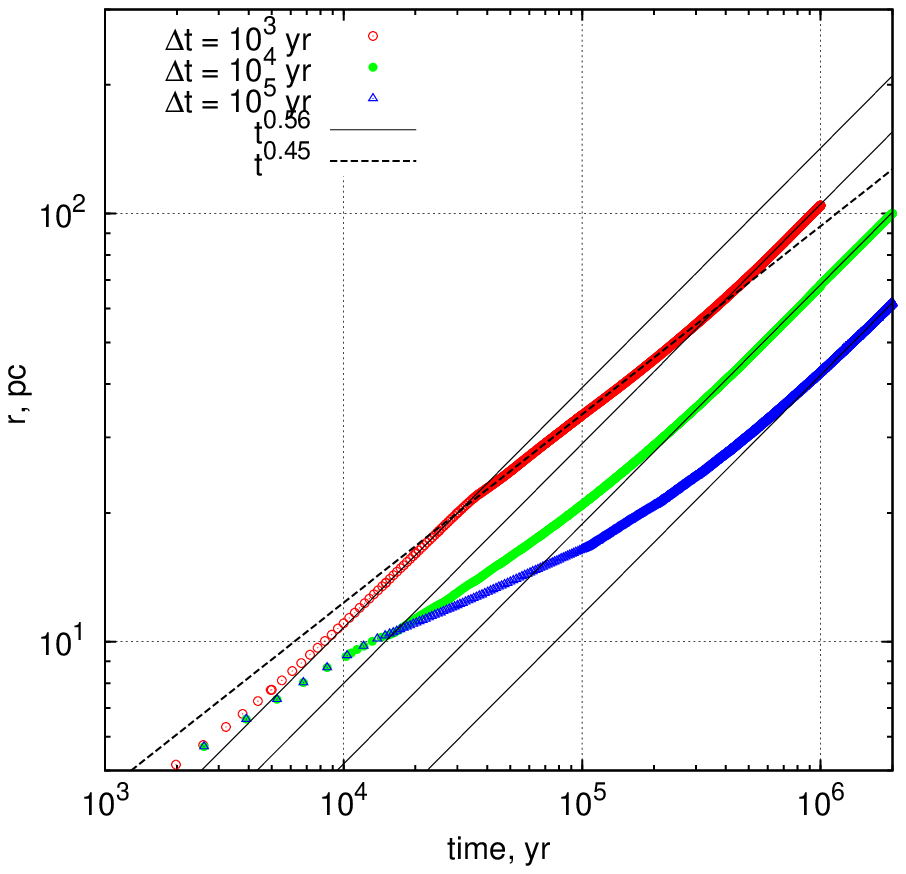}
\caption{
Time dependence of the {\it effective} radius of the collective bubble formed by multiple SNe exploded sequentially in the centre with a constant rate: 1 SN per $10^3, 10^4$ and $10^5$~yr$^{-1}$ are depicted by red, green and blue symbols. The ambient density is 1~cm$^{-3}$ (left panel) and 10~cm$^{-3}$ (right panel). The {\it effective} radius is calculated as a cubic root of the volume of the bubble.
}
\label{fig-r0sn-rad}
\end{figure*} 
%%%%%%%%%%%%%%%%%%%%%%%%%%%%%%%%%%%%%%%%%%%%%%%%%%%%%%%%%%%%%%%%%%%%%

Let us now consider another simple case -- centred SNe explosions, i.e., multiple SNe exploding sequentially at the same point. This commonly used model is considered reasonable since the typical size of a progenitor stellar cluster is much smaller than the size of the remnant produced by the explosion. A detailed description of such a remnant has been performed within a 1D approach \citep[e.g.,][]{roy14}, which allowed for a high spatial resolution, homogeneously spread from the central region to the shell. However, hydrodynamic (Rayleigh-Taylor) and thermal instabilities can influence the dynamics of the shell evolution by enhancing radiation losses. Such an enhancement is mainly due to an increase of the remnant surface by growing `tongues' driven by Rayleigh-Taylor instability \citep{korolev}. This motivates us to describe the centred SNe explosions in 3D. Note though that the resolution we reach in our simulations (0.5 pc) is not quite sufficient for discriminate the clumps driven by thermal instability (the corresponding sizes are of $\sim 0.1$~pc), and may, in principle, suffer from numerical instabilities. This problem is well known for multi-dimensional simulations \citep[e.g.,][]{blinn16}.

%%%%%%%%%%%%%%%%%%%%%%%%%%%%%%%%%%%%%%%%%%%%%%%%%%%%%%%%%%%%%%%%%%%%%
\begin{figure*}
\center 
\includegraphics[width=8cm]{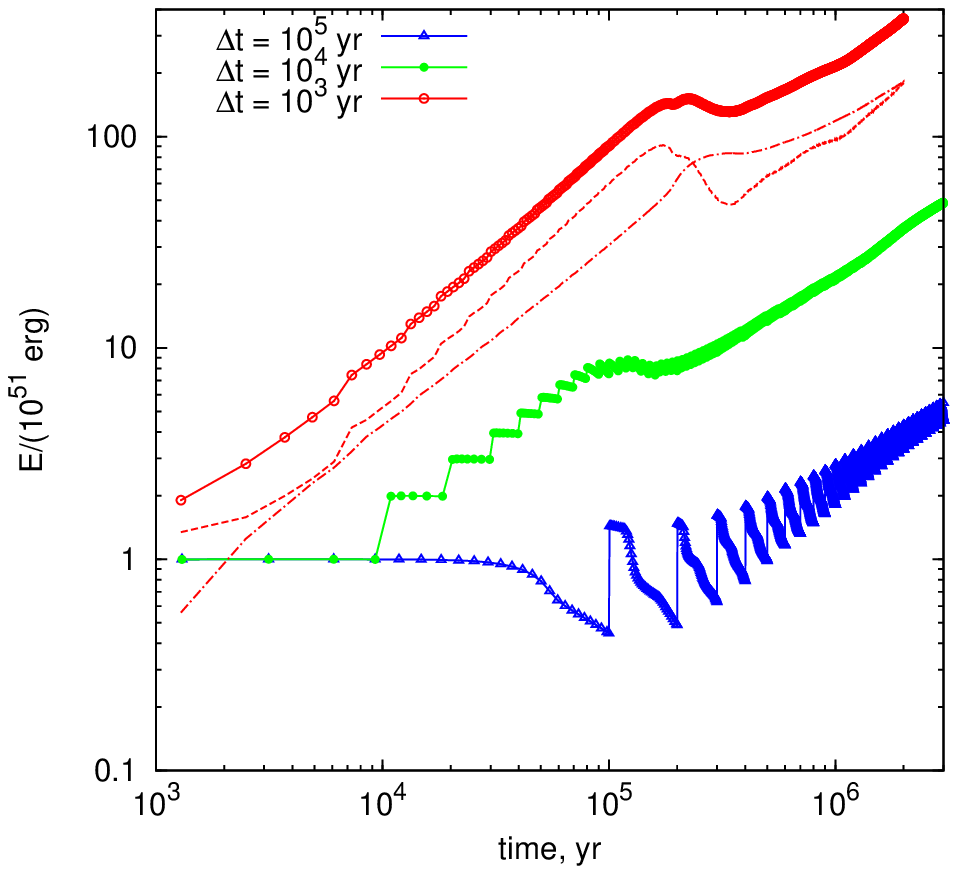}
\includegraphics[width=8cm]{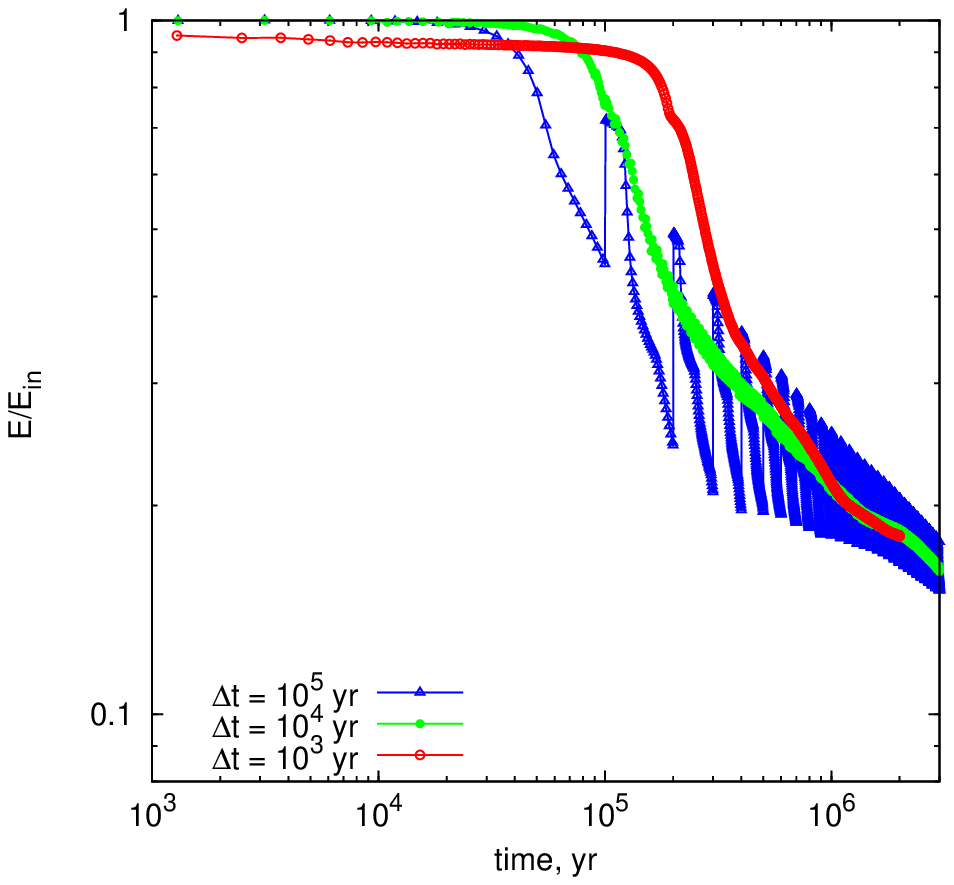}
\break
\includegraphics[width=8cm]{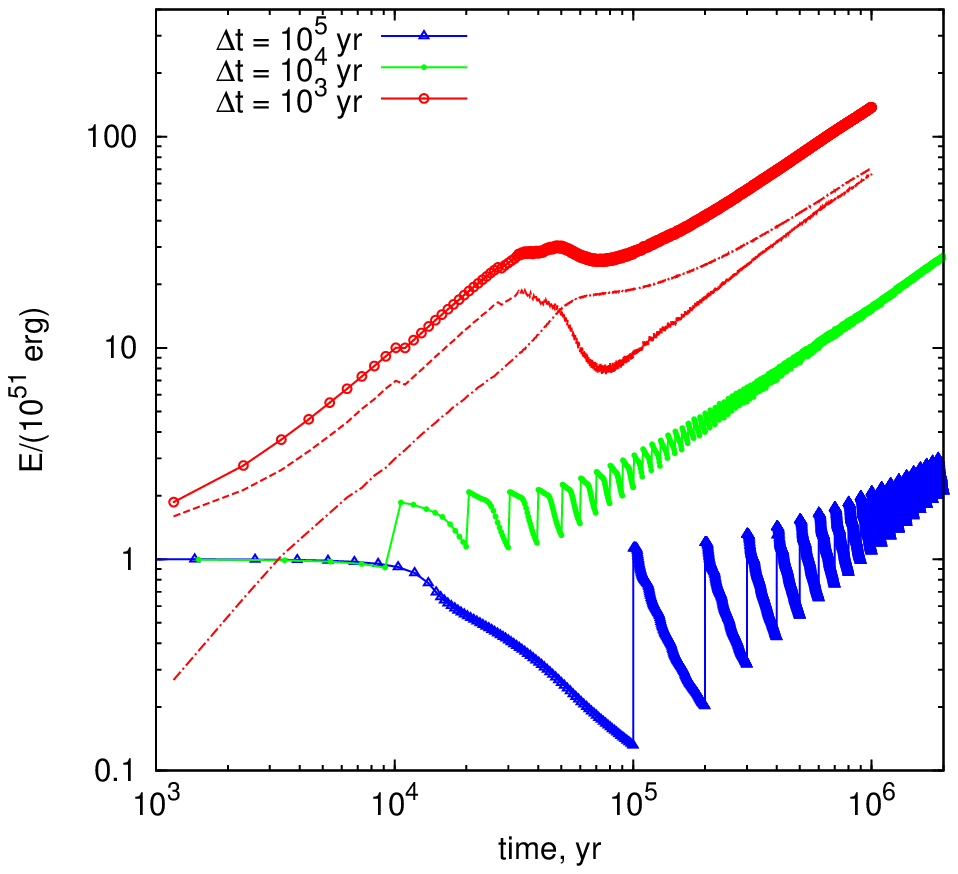}
\includegraphics[width=8cm]{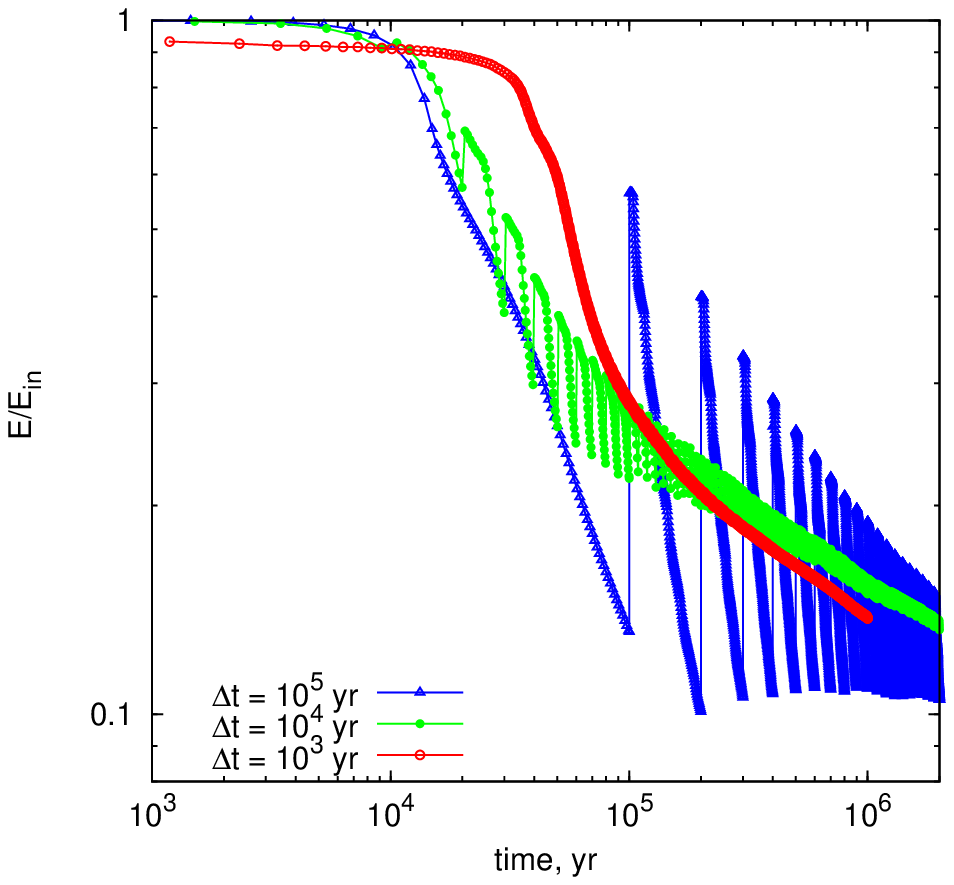}
\caption{
The total energy (left column) and ratio of the total energy to the injected one (right column) of the collective bubble formed by multiple SN explosions in the centre with rate 1 SN per $10^3, 10^4$ and $10^5$~yr$^{-1}$ depicted by red, green and blue symbols. The ambient density equals 1~cm$^{-3}$ (upper panels) and 10~cm$^{-3}$ (lower panels). 
}
\label{fig-r0sn-ene}
\end{figure*} 
%%%%%%%%%%%%%%%%%%%%%%%%%%%%%%%%%%%%%%%%%%%%%%%%%%%%%%%%%%%%%%%%%%%%%

It is worth noting that the time delay between subsequent explosions is an additional characteristic time scale in the problem and its ratio to the cooling time may be an important factor. Indeed, when the time gap between two subsequent explosions $\Delta t$ is shorter than the cooling time $t_c$, the energy supply into the shell can partly replenish the radiative losses and make the expansion more rapid, as seen in the left panel of Figure~\ref{fig-r0sn-rad}. However, as time elapses, radiation losses grow as $\propto R^3$. Ultimately the shell settles on to the wind regime, with radii greater in those cases when $\Delta t/t_c<1$, as seen in Figure~\ref{fig-r0sn-rad} where the  {\it effective} bubble radii driven by collective SN explosions located in the centre are shown. The corresponding cooling time $t_c \sim kT/\Lambda n \sim 10^4 n^{-1}$~yr for the solar metallicity ($\Lambda(T\sim 10^6{\rm K}) \sim 10^{-22}$~erg~s$^{-1}$~cm$^3$). 

This figure shows the results for models with the rates 1 SN per $10^3, 10^4$ and $10^5$~yr in the ISM of ambient density $n=1$ cm$^{-3}$ (left panel), and $n=10$ cm$^{-3}$ (right panel); radiative cooling times are $t_c\sim 10^4$ yr in the left panel and $t_c\sim 10^3$ yr in the right panel, depicted by red, green and blue symbols. It is readily seen that for a delay of $\Delta t = 10^3$~yr, the shell radius initially grows as $\sim t^{0.57}$, close to the wind mode \citep{avedisova,castor}. Initial shell dynamics in models with a longer delay shows a pressure-dominated mode $\sim t^{0.45}$ (see Appendix B), which asymptotically turns to the wind solution. 

Ejecta from successive SN exploding into the already existing hot low-density bubble from earlier SNe expand without losing energy. Its initial dynamics is close to that of the free expansion solution, which evolves to the adiabatic mode (see Appendix~A and Figure~\ref{fig-radevol1sn}). Eventually, the ejecta merges with the shell produced by preceding explosions and transfer to it all their energy.   

Figure~\ref{fig-r0sn-ene} shows the time dependence of the total energy (left-side panels) and the remaining fraction of energy $f_E=E/E_{in}$ (right-side panels) in a remnant from a collective action of sequential SNe centred explosions in two different environments: in a low ($n=1$ cm$^{-3}$, left panel) and high ($n=10$ cm$^{-3}$, right panel) density ambient gas. Several generic features are observed in Fig.~\ref{fig-r0sn-ene} and from a comparison of Figs~\ref{fig-r0sn-rad} and~\ref{fig-r0sn-ene}: {\it i)} a nearly order of magnitude decrease of the characteristic cooling time with the increase in ambient density,  {\it ii)} during radiative stages, the remaining energy fraction decreases approximately as $\propto R^{-2}$ -- much slower than that inferred from 1D simulations of a single SN remnant \citep{cox72}, {\it iii)} remnants with less frequent SNe explosions enter the radiative stage, i.e. show decreasing energy fraction $f_E$, earlier than that with a higher SN rate, {\it iv)} in spite of the fact that at the beginning of the radiative regime the energy drop is larger for the remnants in a denser environment, the asymptotic behaviour of the remaining energy fraction $f_E$ flattens roughly as $R^{-1.8}$, and depends fairly weakly on the ambient density (nearly as $\propto n^{0.2}$).

%----------------------- Section 4 -------------------------------

\section{Clustered SNe}

%%%%%%%%%%%%%%%%%%%%%%%%%%%%%%%%%%%%%%%%%%%%%%%%%%%%%%%%%%
\begin{figure*}
\center
\includegraphics[width=8.5cm]{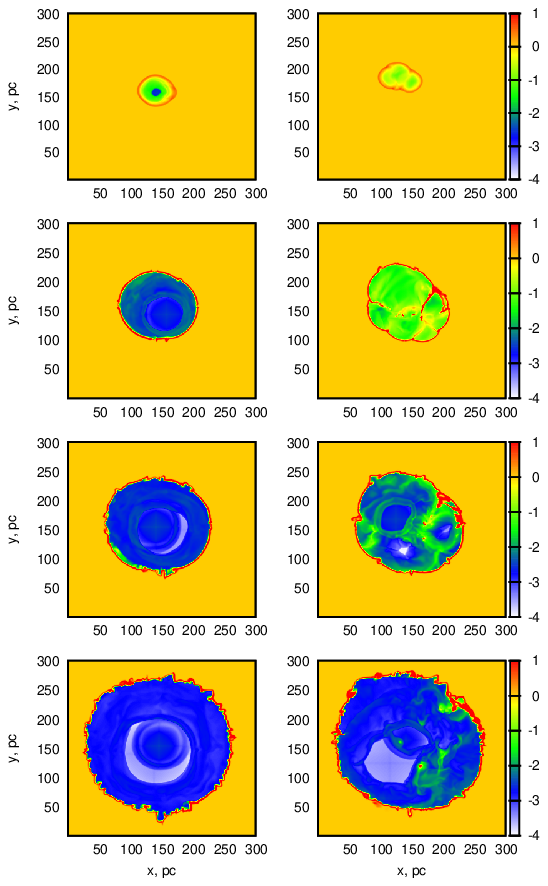}
\includegraphics[width=8.5cm]{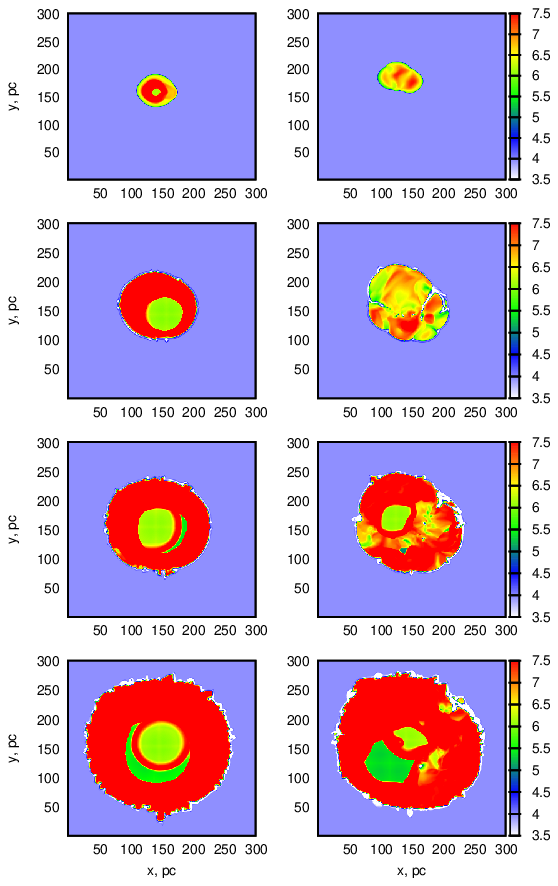}
\caption{
The density (left group of panels) and temperature (right group) slices at $z = 150$~pc for multiple SN explosions in a cluster with volume rate $10^{-9}$~pc$^{-3}$ yr$^{-1}$ (left column in the group) and $4 \times 10^{-11}$~pc$^{-3}$ yr$^{-1}$ (right column in the group) at time moments $5 \times 10^4$, $2.5 \times 10^5$, $5\times 10^5$, $10^6$~yr (from top to bottom). The ambient density equals 1~cm$^{-3}$. Multiple shells and walls are seen in bubbles formed by collective explosions with lower rate for $4 \times 10^{-11}$~pc$^{-3}$ yr$^{-1}$ at times $2.5 \times 10^5$ and $5\times 10^5$~yr. 
}
\label{fig-multisn-maps}
\end{figure*}
%%%%%%%%%%%%%%%%%%%%%%%%%%%%%%%%%%%%%%%%%%%%%%%%%%%%%%%%%%%

Next we consider the effect of having the SNe explode at different (random) locations within a cluster of radius $r_c$.
Figure~\ref{fig-multisn-maps} presents the evolution of density (left group of panels) and temperature (right group of panels) in a bubble growing under the action of multiple SN explosions assembled in a cluster.  The two models with a rate $\nu_{\rm sn}=10^{-9}$~pc$^{-3}$ yr$^{-1}$ (left column in the group) and $4 \times 10^{-11}$~pc$^{-3}$ yr$^{-1}$ (right column in the group) as seen in the plane $z = 150$~pc are shown. In the first case the SNe are spread in a volume of radius $r_c=30$ pc, and in the second, over a volume of $r_c=90$ pc. Throughout the rest of the paper we fix the time lag between subsequent SN explosions to be $\Delta t=10^4$ yr, and connect the variations of the explosion rate $\nu_{\rm sn}$ with variations of the cluster radius $r_c$. The accepted values of $\Delta t$ and $r_c$ reasonably fit real OB-associations. 

In the case of higher SN rate, the collective bubble is already formed by the time $t=5\times 10^4$~yr (top panel), whereas for lower rate, individual remnants still continue to merge until $t=5\times 10^5$~yr (third panel). Therefore bubbles growing under collective explosions with a lower rate remain more irregular on longer time scales with dense walls and broken shells inside the bubble. These fragmentary structures represent either parts of the merged shells or shells from older explosions broken by younger ones. These features are transients with characteristic time scale $\Delta t\simlt (\nu_{\rm sn}r_c^3)^{-1}\sim 10^{4}$ yr. At the same time, signatures of  separate SNe explosions within the bubble can be clearly found in the form of nearly  spherical rings with higher density and temperature, particularly for the higher rate of SNe. A fraction of the gas inside these remnants from individual explosions becomes cool due to adiabatic expansion.  

It is readily seen that the overall dynamics can be described qualitatively in terms of the total number of SNe occurred in the cluster volume with $r_c$ within the cooling time $t_c$, $N_{\rm sn,c}=4\pi\nu_{\rm sn}r_c^3t_c/3$: the lower the number $N_{\rm sn,c}$, the more non-homogeneous and fragmentary is the density and temperature distribution within the superbubble and the larger fraction of energy is lost radiatively (see discussion in the next section).

%%%%%%%%%%%%%%%%%%%%%%%%%%%%%%%%%%%%%%%%%%%%%%%%%%%%%%%%%%%%%%%%%%%%%
\begin{figure*}
\center 
\includegraphics[width=8cm]{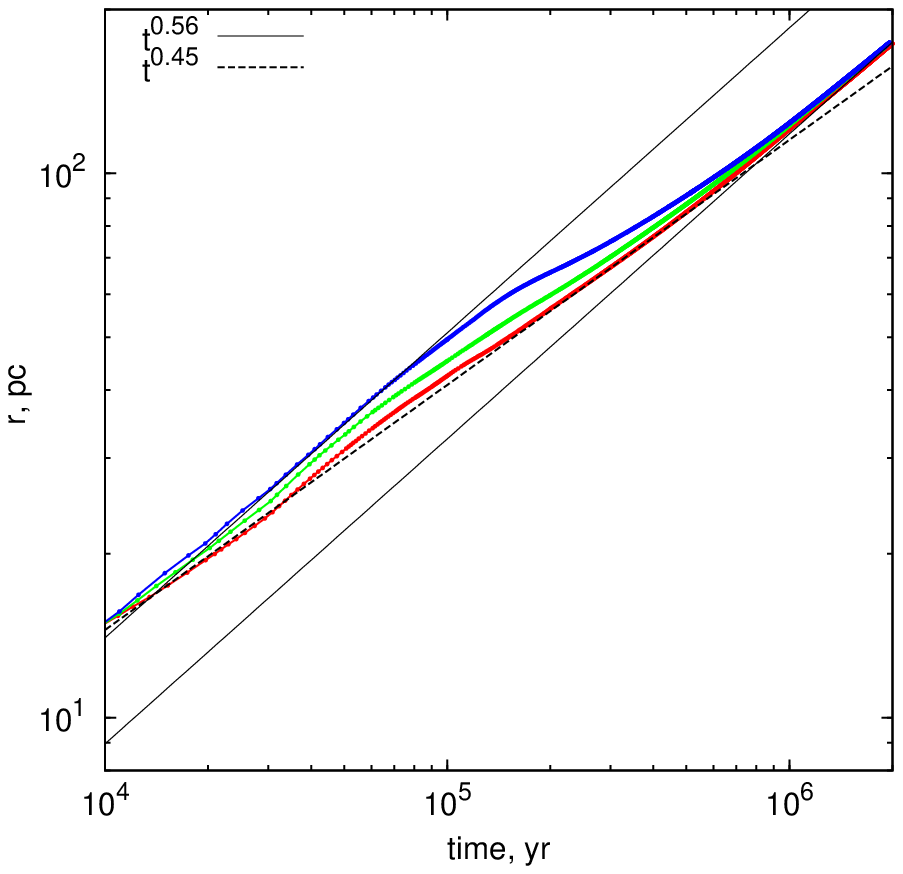}
\includegraphics[width=8cm]{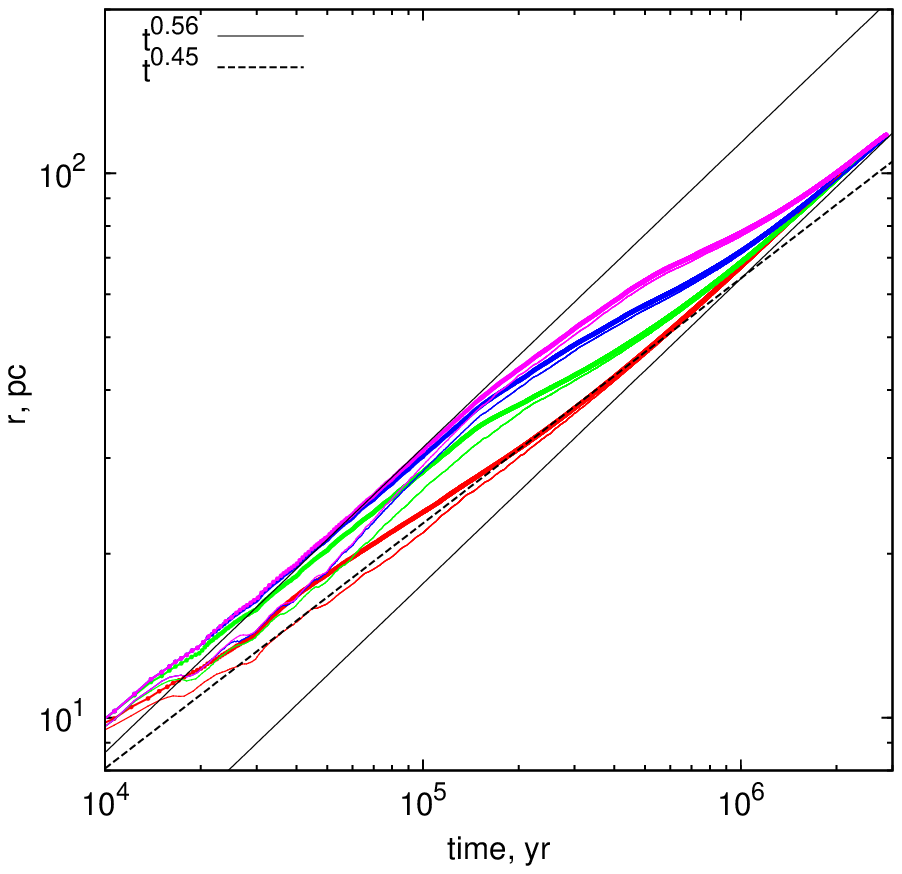}
\caption{
Evolution of the {\it effective} radius $\langle R(t)\rangle=V^{1/3}$ of the (super)bubble formed by clustered multiple SNe with the rate $10^{-9}, 1.2 \times 10^{-10}, 4 \times 10^{-11}$pc$^{-3}$ yr$^{-1}$ depicted by red, green and blue { (from bottom to top)} lines in both panels and $1.5 \times 10^{-11}$pc$^{-3}$ yr$^{-1}$ at the right panel by purple (upper) line. Left panel is for the ambient density 1~cm$^{-3}$, while right panel corresponds to 10~cm$^{-3}$.Thin lines on the right panel show the {\it effective} radius determined via extraction of the SN bubble by temperature difference -- the two techniques give 5--10\% differences in the {\it effective} radius at early and intermediate stages $t\simlt 10^5$ yr. 
}
\label{fig-multisn-rad}
\end{figure*} 
%%%%%%%%%%%%%%%%%%%%%%%%%%%%%%%%%%%%%%%%%%%%%%%%%%%%%%%%%%%%%%%%%%%%%

Figure~\ref{fig-multisn-rad} presents the evolution of the {\it effective} radius $\langle R(t)\rangle=V^{1/3}$ of the (super)bubble from multiple clustered SN for a set of explosion rates. In order to extract the volume occupied by remnants we fiducially applied the following technique: a single cell from the whole computational zone was counted as belonging to the remnant provided its velocity $\langle v_i^2\rangle^{1/2}=\sqrt{v_{i,x}^2+v_{i,y}^2+v_{i,z}^2}$ exceeds 1~km~s$^{-1}$ -- the corresponding results are shown by thick lines on Figure~\ref{fig-multisn-maps}. As an alternative, we also applied a technique based on the temperature differences: a cell was assigned to the superbubble if gas temperature within the cell deviates from the unperturbed ambient temperature $T=10^4$ K by $\Delta T\simlt 100$~K -- the corresponding {\it effective} radii are shown by thin lines in the right panel on Figure~\ref{fig-multisn-maps}.

For the highest rate $\nu_{\rm sn} = 10^{-9}$pc$^{-3}$ yr$^{-1}$ (red lines in left panel of Figure~\ref{fig-multisn-rad}), SN remnants merge at very early stages (see upper left panel in Figure~\ref{fig-multisn-maps}), such that in the low-density environment, $n=1$~cm$^{-3}$, several SNe explode in a single remnant before it enters the radiative cooling stage. As a result, the collective bubble transforms into a wind-like regime as manifested on Figure~\ref{fig-multisn-rad} at $t\sim 3\times 10^4$~yr as $r(t)\sim a_1t^{0.56}$ (with the exponent 0.56 reasonably close to the standard exponent 0.6 for a wind regime). However, quite soon radiative losses come into play and the expansion slows down to a slower scaling of $t^{0.45}$ and  eventually sets again on to the same power-law $\sim a_2t^{0.56}$ with $a_2<a_1$. The factor {$a_2$ is immediately} connected to the supernova explosion energy and their rate as {$a_2\simeq (L/\rho)^{1/5}$} with the effective mechanical luminosity $L=0.3 E/\Delta t$ (see Appendix~C). 

For lower rates $\nu_{\rm sn}\leq 1.2\times 10^{-10}$pc$^{-3}$ yr$^{-1}$ (green and blue lines in the left panel of Figure~\ref{fig-multisn-rad}), several initial SNe expand in isolation with the {\it effective} radius $r\sim t^{0.56}$ as described in Section~2.1. After $t\sim 3\times 10^4$~yr, the remnant turns to the radiative stage and evolves further as in the previous case of $\nu_{\rm sn} = 10^{-9}$pc$^{-3}$ yr$^{-1}$. It is worth pointing out a seemingly unusual increase in the {\it effective} radius with decreasing explosion rate as seen in Figure~\ref{fig-multisn-rad}. It can be understood in terms of the superposition of adiabatic remnants in limiting cases: in the low rate end, the total volume is assembled from several nearly isolated remnants $\langle R(t)\rangle\sim [\sum_i R(t-t_i)^3]^{1/3}\propto t^{2/5+1/3}$, while in the high rate end, different explosions add energy approximately into the same remnant $\langle R(t)\rangle\sim [\sum_i  E_i(t-t_i)^2/\rho]^{1/5}\propto t^{3/5}$. However, at later stages when the remnants fully merge, the differences between the time dependences of {\it effective} radii vanish. 

The evolution of a growing superbubble in gas with higher density is mostly similar to that described above with only minor deviations: they enter the radiative regime very early at around $3\times 10^3$ yr, such that at initial stages the {\it effective} radii $\langle R(t)\rangle$ show a power-low expansion $\sim a_1t^{0.56}$ with slightly irregular variations due to fast cooling. Another  feature is that the remnants radiatively lose more energy, and as a result their asymptotic power-law expansion $\sim a_2t^{0.56}$ lies lower than in models with $n=1$ cm$^{-3}$, i.e. the difference between $a_1$ and $a_2$ in models with $n=10$ cm$^{-3}$ is larger than at lower densities. As seen from comparison of the expansion laws in Figure~\ref{fig-multisn-rad} {$a_2\simeq 0.6a_1$ at $n=1$ cm$^{-3}$ while at $n=10$ cm$^{-3}$ $a_2\simeq 0.5a_1$.} 

%%%%%%%%%%%%%%%%%%%%%%%%%%%%%%%%%%%%%%%%%%%%%%%%%%%%%%%%%%%%%%%%%%%%%
\begin{figure*}
\center 
\includegraphics[width=8cm]{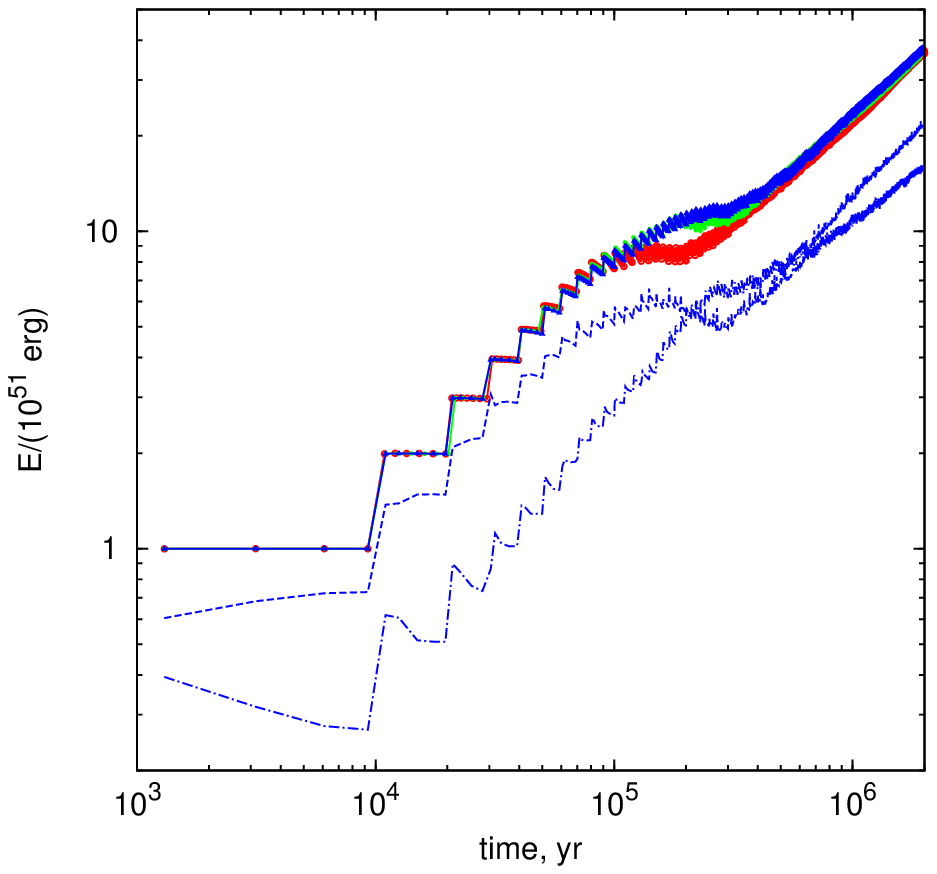}
\includegraphics[width=8cm]{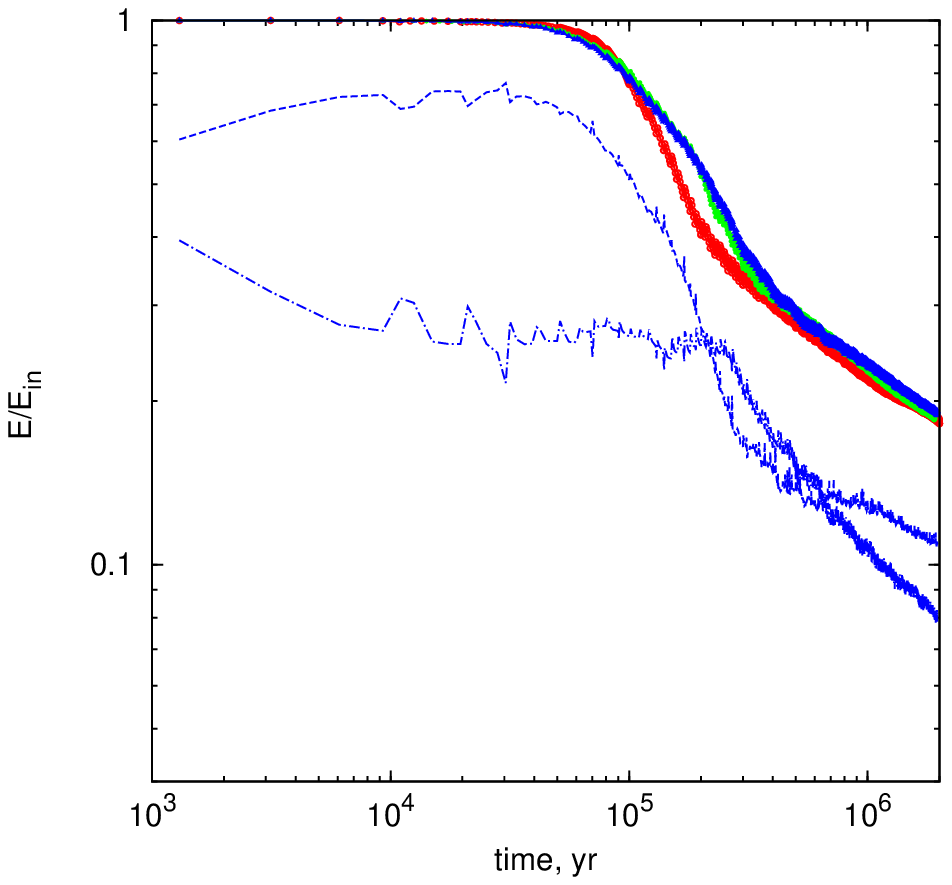}
\break
\includegraphics[width=8cm]{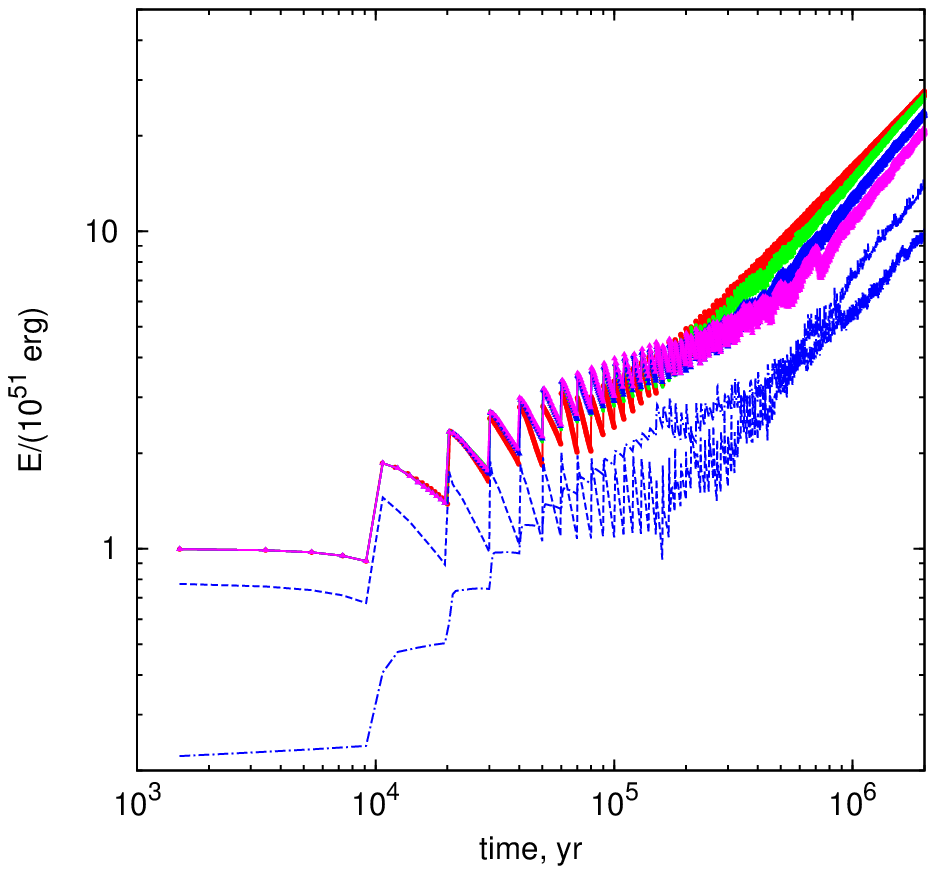}
\includegraphics[width=8cm]{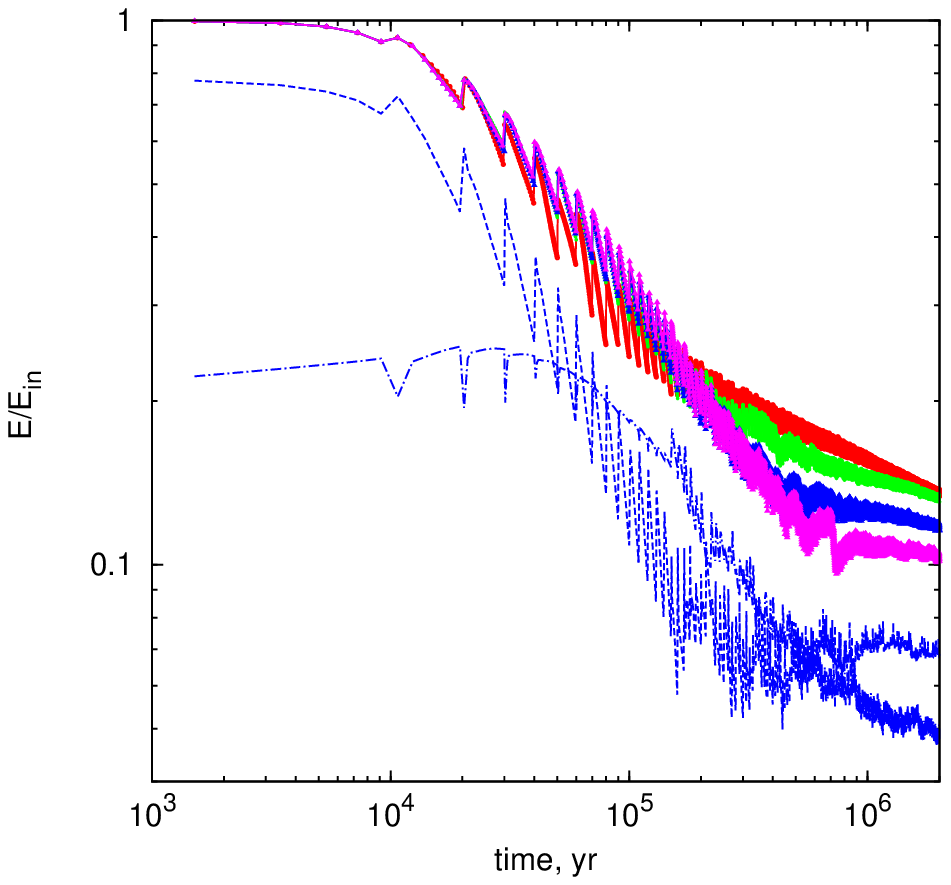}
\caption{
The total energy (left column) and ratio of the total energy to the injected one (right column) of the collective bubble formed by multiple SN explosions in a cluster with volume rate $10^{-9}, 1.2 \times 10^{-10}, 4 \times 10^{-11}$pc$^{-3}$ yr$^{-1}$ depicted by red, green and blue lines in both panels and $1.5 \times 10^{-11}$pc$^{-3}$ yr$^{-1}$ at bottom panel by purple lines. The ambient density equals 1~cm$^{-3}$ (upper panels) and 10~cm$^{-3}$ (lower panels). 
For illustration thermal (dashed line) and kinetic (dash-dotted line) energies (left panels) and their fractions (right panels)  are shown for the rate $4 \times 10^{-11}$pc$^{-3}$ yr$^{-1}$.
}
\label{fig-multisn-ene}
\end{figure*} 
%%%%%%%%%%%%%%%%%%%%%%%%%%%%%%%%%%%%%%%%%%%%%%%%%%%%%%%%%%%%%%%%%%%%%

To illustrate the expansion laws, we show in Figure~\ref{fig-multisn-ene} the total energy (left column) of the remnant including thermal energy and kinetic energy of the bubble and the shells and their fragments. All characteristic features seen in models with centred explosions are present here and reflected in the dependences $\langle r(t)\rangle$. It is seen in particular that the overall energy drop is larger at higher density (right lower panel), which causes a stronger decrease in the asymptotic behaviour of the {\it effective} radius. 

%%%%%%%%%%%%%%%%%%%%%%%%%%%%%%%%%%%%%%%%%%%%%%%%%%%%%%%%%%%%%%%%%%%%%
%\begin{figure*}
%\center 
%\includegraphics[width=8cm]{mas-r30-90-n1.eps}
%\includegraphics[width=8cm]{mas-r30-120-n10.eps}
%\caption{
%The total mass of the collective bubble formed by multiple SN explosions in a cluster with volume rate $10^{-9}, 1.2 \times 10^{-10}, 4 \times 10^{-11}$pc$^{-3}$ yr$^{-1}$ depicted by red, green and blue symbols in both panels and $1.5 \times 10^{-11}$pc$^{-3}$ yr$^{-1}$ at bottom panel by purple symbols. The ambient density equals 1~cm$^{-3}$ (lower panel) and 10~cm$^{-3}$ (right panel).
%}
%\label{fig-multisn-mas}
%\end{figure*} 
%%%%%%%%%%%%%%%%%%%%%%%%%%%%%%%%%%%%%%%%%%%%%%%%%%%%%%%%%%%%%%%%%%%%%

%%% Figure~\ref{fig-multisn-mas} presents the evolution of total mass of the collective bubble formed by multiple SN explosions in a cluster for the same volume rates as in Figure~\ref{fig-multisn-rad}. It is expected that the mass of the collective bubble evolves in similar manner as its radius. At late time the mass grows almost as $\sim t^{1.5}$

%%%%%%%%%%%%%%%%%%%%%%%%%%%%%%%%%%%%%%%%%%%%%%%%%%%%%%%%%%%%%%%%%%%%%
\begin{figure*}
\center 
\includegraphics[width=8cm]{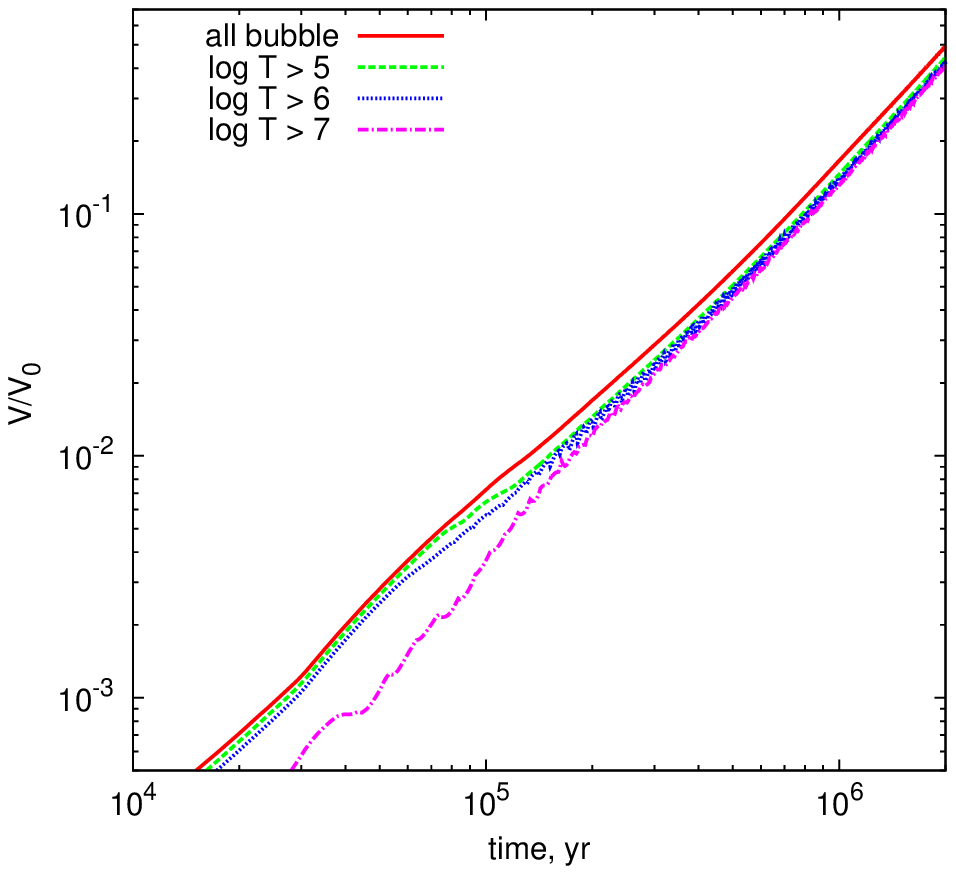}
\includegraphics[width=8cm]{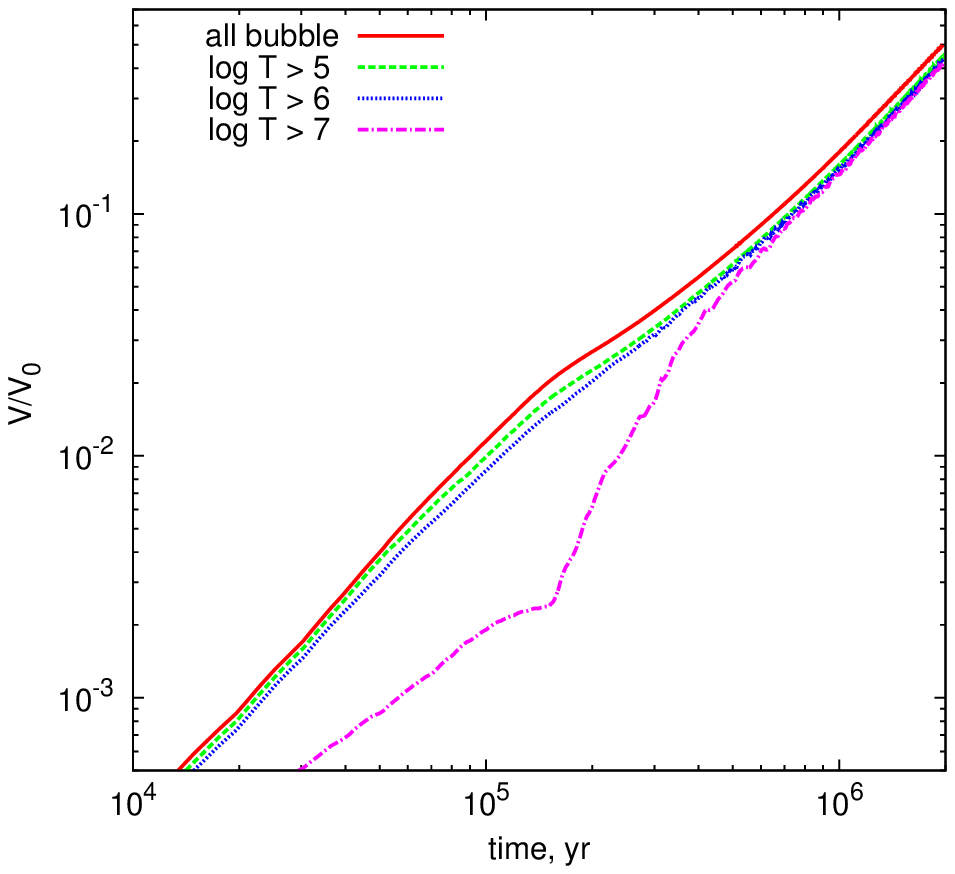}
\caption{
Volume filling factors of gas with different temperatures inside the collective bubble formed by multiple SN explosions in a cluster with volume rate $10^{-9}$pc$^{-3}$ yr$^{-1}$ (left panel) and  $4 \times 10^{-11}$pc$^{-3}$ yr$^{-1}$ (right panel). The ambient density equals 1~cm$^{-3}$. The total volume of the computational domain equals to $V_0$.
}
\label{fig-multisn-volt}
\end{figure*} 
%%%%%%%%%%%%%%%%%%%%%%%%%%%%%%%%%%%%%%%%%%%%%%%%%%%%%%%%%%%%%%%%%%%%%

The volume filling factors of gas with different temperatures (temperature higher than a given level) are given in Figure~\ref{fig-multisn-volt}. We notice a clear difference in the evolution of the hottest gas: its fraction is much less than unity and that of the colder components, and remains so until the remnants from different SNe overlap and the hot interiors merge.

%----------------------- Section 5 -------------------------------

\section{Observational consequences} 

\subsection{Kinematics}

Radius of a wind blown bubble evolves as $r\simeq 0.76 (L/\rho)^{1/5}t^{3/5}$, where $L = \dot M v^2/2$ is the mechanical luminosity \citep{castor}. In simulations of star-forming galaxies and the interpretation of observational data, the mechanical luminosity is commonly assumed to be the ratio of the total energy of all SNe to the duration of a starburst: $L = N E_{SN}/t$ \citep[e.g.,][]{mccray88}. In case of a constant SN rate, $L = E_{SN}/\Delta t$, where $E_{SN}$ is the energy of a single SN, $\Delta t$ is the time delay between  successive SNe. Accordingly, the radius of a superbubble from this SNe ensemble can be expressed as 
\be
r \simeq 160 \left({10^4 {\rm yr} \over \Delta t } {1 {\rm cm^{-3}} \over n }\right)^{1/5} \left( {t\over 10^6 {\rm yr}}\right)^{3/5} \ {\rm pc}. 
\label{rwind}%0.01\%.
\ee
It coincides with the previously described numerical models with $\Delta t = 10^3$~yr during the early evolution, before the  radiative stage begins at $t\sim 1.5\times 10^5$~yr for $n=1$~cm$^{-3}$ and $t\sim 3\times 10^4$~yr for $n=10$~cm$^{-3}$ (Figure~\ref{fig-r0sn-rad}). 

After passing through the intermediate phase, the superbubble eventually settles on to a new power-law wind asymptote but with the radius being a factor of 1.4 ($n=1$ cm$^{-3}$) to 2 ($n=10$ cm$^{-3}$) smaller than given by (\ref{rwind}). This leads to an underestimate of the mechanical luminosity $L$ by about $\sim 1.4^{5} \simeq 5.4$ to $\sim 2^5=32$ when $L$ is estimated  from the directly observable quantities: radius $R$, expansion velocity $v$ and the shell mass $M$. 

%dddd

%%%%%%%%%%%%%%%%%%%%%%%%%%%%%%%%%%%%%%%%%%%%%%%%%%%%%%%%%%%%%%%%%%%%%
\begin{figure*}
\center
\includegraphics[width=8cm]{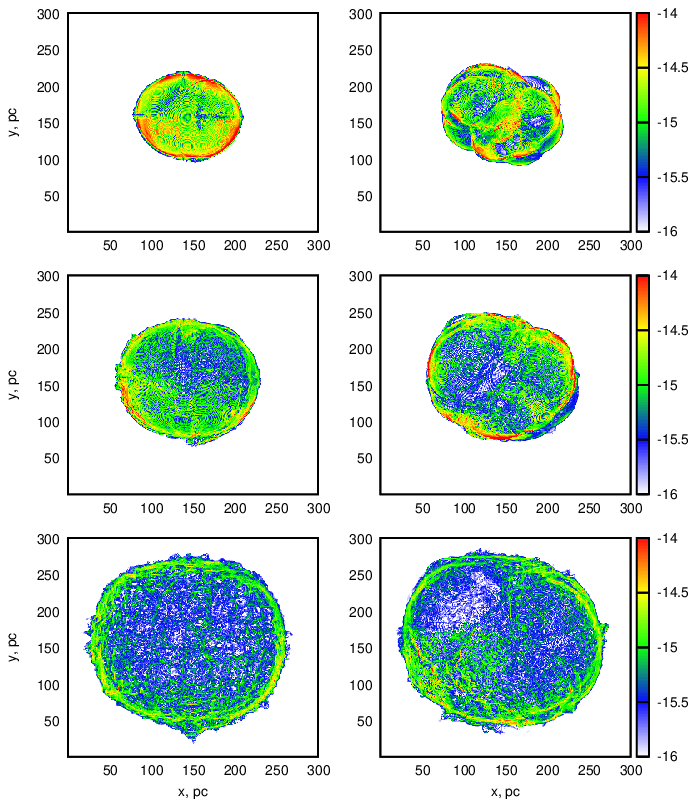}
\includegraphics[width=8cm]{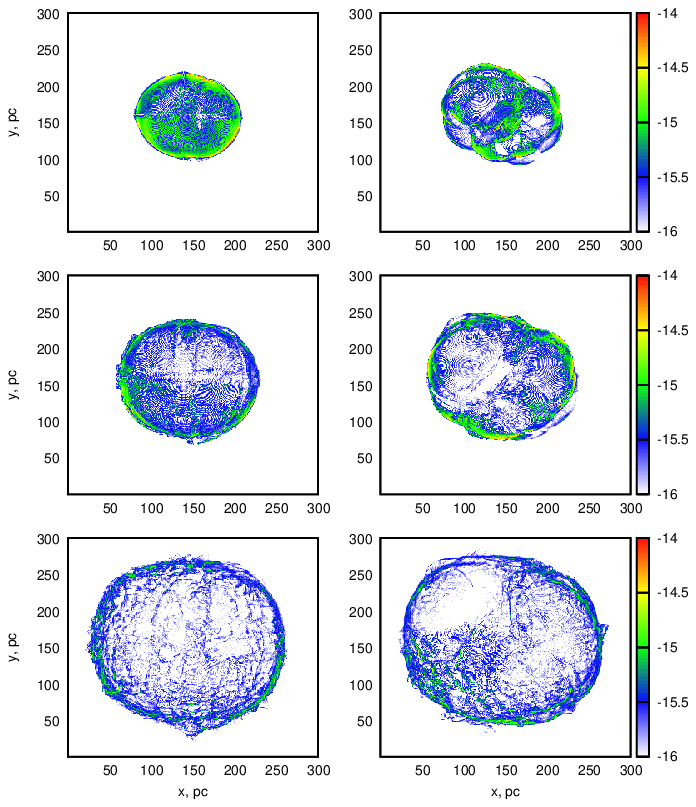}
\caption{
{\it Left panel}: H$\alpha$ surface brightness map (erg cm$^{-2}$ s$^{-1}$ arcsec$^{-2}$ in log scale) of the collective bubble formed by multiple SN explosions in a cluster with volume rate $10^{-9}$pc$^{-3}$ yr$^{-1}$ (left column) and  $4 \times 10^{-11}$pc$^{-3}$ yr$^{-1}$ (right column) at time moments $2.5 \times 10^5$, $5\times 10^5$, $10^6$~yr (from top to bottom). {\it Right panel}:  H$\beta$ surface brightness map (erg~cm$^{-2}$~s$^{-1}$~arcsec$^{-2}$ in log scale) for the same models as in the left panel. The ambient density equals 1~cm$^{-3}$. 
}
\label{fig-multisn-ha-map}
\end{figure*} 
%%%%%%%%%%%%%%%%%%%%%%%%%%%%%%%%%%%%%%%%%%%%%%%%%%%%%%%%%%%%%%%%%%%%%

\subsection{Emission in Balmer lines}
A useful diagnostic of the superbubble evolution is commonly thought to come from line emission of different elements that probe the physical conditions in gas. Giant holes in the gas distribution in galactic disks of several nearby dwarf galaxies have been studied with the help of H$\alpha$ emission from the gas ionized by Ly-continuum photons produced by the underlying stellar population \citep[e.g.,][]{moiseev07,moiseev12,egorov14,egorov17}. In this environment the spatial H$\alpha$ distribution traces the regions with a higher emission measure $EM=\int n_e^2dl$, and thus reflects variations of gas density and radiation field under an implicit assumption that photoionized gas is kept at $T\simeq 10^4$ K. Under such conditions the {ratio of H$\alpha$ to H$\beta$ intensities} is practically fixed at 2.86 \citep[e.g.,][]{drainebook} provided the dust extinction is weak.

{The situation changes in the case when ionization is dominated by shock waves randomly impinging on a given gas element \citep[see e.g.,][]{raga15}}. Figure~\ref{fig-multisn-ha-map} presents the predicted H$\alpha$ {brightness} maps for superbubbles with SN rate $10^{-9}$pc$^{-3}$ yr$^{-1}$ (left column) and  $4 \times 10^{-11}$pc$^{-3}$ yr$^{-1}$ (right column) at epochs $2.5 \times 10^5$, $5\times 10^5$, $10^6$~yr (top to bottom); {the maps correspond} to the three bottom panels in Figure~\ref{fig-multisn-maps}. It is worth noting here that in contrast with \citet{raga15} we calculated the emission in H$\alpha$ and H$\beta$ lines under the conditions of a non-steady radiatively cooling environment as described above in Section 2. {At earlier times, H$\alpha$ maps are smoother for the higher SN rate, while} those for the lower SN rate reveal copious bright knots and walls inside. It is clear that more frequent SNe sweep the gas from the central part of the bubble outward during the initial explosion episodes, so that every subsequent SN  explodes in a hot low-density environment. Less frequent SNe approach a similar behaviour at later stages when many isolated remnants merge. {Thus, in this case the variations of H$\alpha$ emission are due to variations in density and to some extent in temperature.}

This circumstance is clearly seen in the distribution {function} of the H$\alpha$ {brightness} on Figure~\ref{fig-multisn-ha-histo}; {the distribution is normalized to the total number of cells inside the remnant. The distribution for higher SNe rate is slightly narrower, and over time, shifts  towards a less bright state than that for the lower rate: if at $t=0.25$ Myr the distribution for the bubble with faster SNe peaks at higher brightness $\simeq 10^{-14.7}$~erg~cm$^{-2}$~s$^{-1}$~arcsec$^{-2}$ than for the slower SNe explosions at $\simeq 10^{-15.3}$~erg~cm$^{-2}$~s$^{-1}$~arcsec$^{-2}$, {while} at $t=1$ Myr they both peak at $\simeq 10^{-15.4}$~erg~cm$^{-2}$~s$^{-1}$~arcsec$^{-2}$. Thus, the H$\alpha$ maps in bubble from more frequent SNe explosions ``cool'' faster than those with a lower SN rate. It stems from the fact that frequent SNe explosions empties the interior sweeping gas into the shell where the electrons recombine faster. Note that the distribution functions for H$\beta$ brightness are practically identical to those for H$\alpha$ though slightly shifted towards lower brightnesses. }  In general, the supershells produced by merging remnants from isolated SNe are the brightest regions of the superbubbles. However, for low SN rates, a growing superbubble passes through the intermediate stages when the most prominent features of the merging process -- namely, the fragments of dense overlapping shells and dense clumps --  fill the interior  and  manifest themselves as the brightest spots, as seen in the maps  on Figure~\ref{fig-multisn-ha-map} and in the histogram on Figure~\ref{fig-multisn-ha-histo} {as a wider distribution}. 

%%%%%%%%%%%%%%%%%%%%%%%%%%%%%%%%%%%%%%%%%%%%%%%%%%%%%%%%%%%%%%%%%%%%%
\begin{figure}
\center
\includegraphics[width=8.3cm]{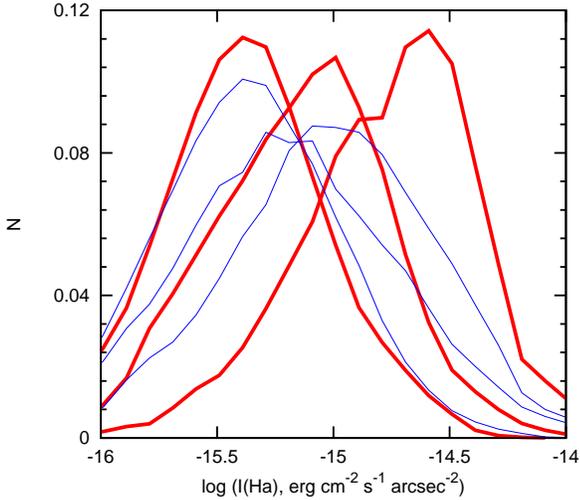}
\caption{
H$\alpha$ surface brightness (in log scale) distributions for the SN rates $10^{-9}$pc$^{-3}$ yr$^{-1}$ (red thick lines) and  $4 \times 10^{-11}$pc$^{-3}$ yr$^{-1}$ (blue thin lines) at time moments {0.25, 0.5 and 1 Myr (right to left)}. The ambient density is 1~cm$^{-3}$. 
}
\label{fig-multisn-ha-histo}
\end{figure} 
%%%%%%%%%%%%%%%%%%%%%%%%%%%%%%%%%%%%%%%%%%%%%%%%%%%%%%%%%%%%%%%%%%%%%

%%%%%%%%%%%%%%%%%%%%%%%%%%%%%%%%%%%%%%%%%%%%%%%%%%%%%%%%%%%%%%%%%%%%%
\begin{figure}
\center
\includegraphics[width=8cm]{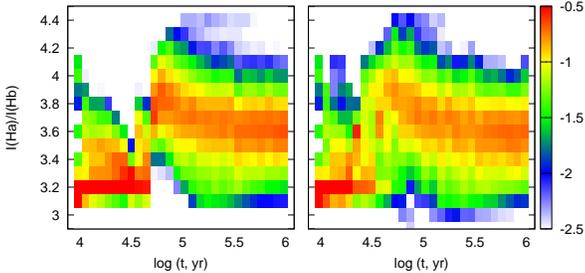}
\caption{
Evolution of the distribution function of the intensity ratio $I({\rm H}\alpha)/I({\rm H}\beta)$ for models shown in Figure~\ref{fig-multisn-ha-map} for the higher $10^{-9}$pc$^{-3}$ yr$^{-1}$ (left panel) and the lower $4 \times 10^{-11}$pc$^{-3}$ yr$^{-1}$ (right panel) SN rates, respectively. The colour bar rightside shows the logarithm fraction of the bubble volume with the intensity ratio in a given range.
}
\label{fig-multisn-ha-hb-ave}
\end{figure} 
%%%%%%%%%%%%%%%%%%%%%%%%%%%%%%%%%%%%%%%%%%%%%%%%%%%%%%%%%%%%%%%%%%%%%

{However, in} practice, the spatial variations of the H$\alpha$ brightness {and their pattern} similar to those shown on Figure~\ref{fig-multisn-ha-map} can hardly be distinguished observationally because of insufficient  resolution: the pixel size in our maps is 1~pc, whereas even nearby galaxies are resolved at {the} best with 5~pc \citep[e.g.,][]{moiseev07,moiseev12}. In order to compare the simulated H$\alpha$ maps to those obtained observationally one would have to degrade the resolution \citep{v15ha}, which would diminish the differences in the patterns of H$\alpha$ brightness between the models shown in  the left and in the right panels of Figure~\ref{fig-multisn-ha-map}. 

The ratio of H$\alpha$ to $H\beta$ intensities would be an additional indicator of the evolutionary status of the superbubble. In the conditions prevailing in the cooling gas in individual remnants and their shells, the fractional ionization stays nearly frozen {because recombination is slower than cooling}. Under such conditions, even rather cold (e.g., $T\simlt 10^2$ K) regions can be luminous in H$\alpha$ and H$\beta$. The ratio ${\rm H}\alpha/{\rm H}\beta$ depends on temperature and under conditions of high fractional ionization ($x>0.5$), it can vary by factor of $\simeq 1.5-2.5$ in the temperature range between $10^3$ K to $10^{5.5}$~K \citep{raga15},  and {thus} can be measured observationally. It is worth noting that in all the models considered here with $\nu_{\rm sn}=(4\times 10^{-11}-10^{-9})$ pc$^{-3}$ yr$^{-1}$, dust extinction along the remnant radius is $A_v\sim (0.1-0.2)\times n^{4/5}(r_{100}t_{\rm Myr})^{3/5}$, where $r_{100}=r_c/(100~{\rm pc})$, $t_{\rm Myr}$ is time elapsed since the first SN in the cluster. This means that for superbubbles grown under a ``normal'' galactic environment with ambient density $n\simlt 1$ cm$^{-3}$, dust extinction cannot affect variations of ${\rm H}\alpha/{\rm H}\beta$ due to evolutionary effects. 

Radiatively cooling gas in the growing bubble stays highly ionized ($x_e\simlt 1$), except in dense parts of the external shell and fragments of shells from overlapping remnants in the bubble. At the same time, temperature can considerably vary from cell to cell  resulting in the variation of intensities of H$\alpha$ and H$\beta$ lines throughout the whole bubble.  It differs from the case of a predominantly photoionized gas where the ratio is practically fixed at 2.86. 

Figure~\ref{fig-multisn-ha-hb-ave} shows {the distribution function of H$\alpha$ to H$\beta$ ratio for the two superbubbles driven by SNe with the {higher} (left panel) and the {lower} (right panel) rates  over the time range} from $10^4$ to $10^6$ yr. 
The colour bar in the right shows the logarithm of the fraction of volume occupied, such that the distribution function varies from $f\sim 3\times 10^{-3}$ to $\sim 0.3$, peaking at ${\rm H}\alpha/{\rm H}\beta=3.6$ practically independent of the SN rate. It is readily seen, that the ratio H$\alpha$ to H$\beta$, for the interior of superbubbles differs from the ``standard'' value of photoionized gas ${\rm H}\alpha/{\rm H}\beta=2.86$ at $T\simeq 10^4$ K. {This difference, however, is  small to be distinguished observationally. However, the spread of {the ${\rm H}\alpha/{\rm H}\beta$ ratio} can reach upto factor 1.5, and this result may be important in distinguishing photoionized regions of a superbubble from those where ionization is predominantly collisional due to multiple randomly impinging shock waves. As mentioned above, the dust extinction in this context of superbubble dynamics%under conditions when superbubble dynamics is concerned 
has small effects on the variations of {the ${\rm H}\alpha/{\rm H}\beta$ ratio}. For instance, in Holmberg II the total extinction in H$\alpha$ emitting gas is only $A_v\simeq 0.1$, assuming solar metallicity} \citep[e.g.,][]{egorov17}. 

\subsection{X-ray maps} 
A simple inspection of Figure \ref{fig-multisn-volt} confirms that a considerable fraction of the remnant from collective SNe is occupied by hot gas. Thus one can expect not only optical emission in hydrogen Balmer recombination lines from the cold remnant shell, but also X-ray photons from its hot interior. This issue is important because the gas density inside the remnants is low. It is clearly seen from Fig. \ref{fig-multisn-maps} where for both models with the lower and higher SN rates, the gas density inside the growing bubble at late stages ($t>1$ Myr) is mostly lower than  $n\simlt 0.01$ cm$^{-3}$, such that pressure within it is a few $\times 10^4$ K cm$^{-3}$. 

%%%%%%%%%%%%%%%%%%%%%%%%%%%%%%%%%%%%%%%%%%%%%%%%%%%%%%%%%%%%%%%%%%%%%
\begin{figure*}
\center
\includegraphics[width=16cm]{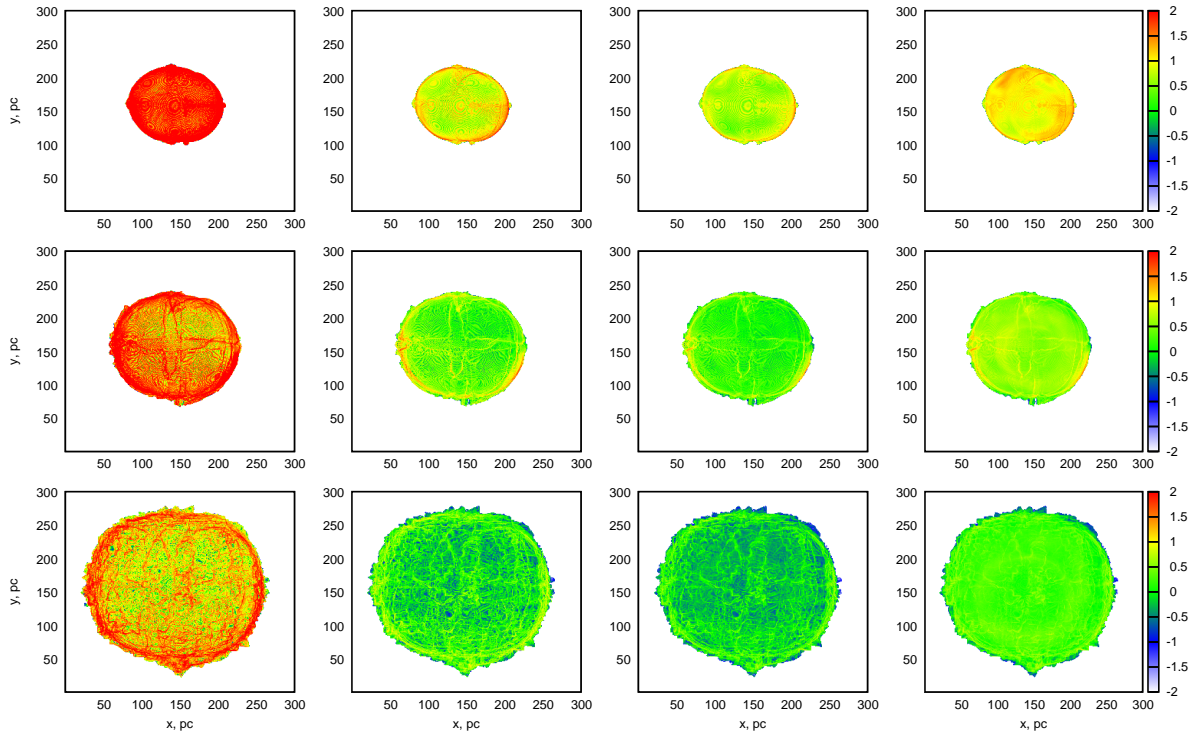}
\caption{
X-ray surface brightness of the bubble in different energy bends (left to right; $0.1-07$, $0.7-1.2$, $1.2-2.1$ and $1.6-8.3$ keV) and of different age of the remnant (top to bottom: 0.25, 0.5 and 1 Myr) for  the bubble formed by SNe explosions with the rate $10^{-9}$pc$^{-3}$ yr$^{-1}$ clustered within 30 pc radius, ambient density is 1~cm$^{-3}$. The units (logarithm scale) in colour bars are keV cm$^{-2}$ s$^{-1}$ sr$^{-1}$.
}
\label{exray_30_evol}
\end{figure*} 
%%%%%%%%%%%%%%%%%%%%%%%%%%%%%%%%%%%%%%%%%%%%%%%%%%%%%%%%%%%%%%%%%%%%%

Figures \ref{exray_30_evol} and \ref{exray-90_evol} demonstrate the spatial distribution of surface X-ray brightness for the remnants for SN rate  $10^{-9}$pc$^{-3}$ yr$^{-1}$ from a cluster of $r_c=30$ pc, and  $4 \times 10^{-11}$pc$^{-3}$ yr$^{-1}$ in a cluster with $r_c=90$ pc, respectively. The columns from left to right depict surface brightness for different energy bands: $0.1 - 0.3$, $0.7 - 1.2$, $1.2-2.1$ and $1.6-8.3$ keV (as in \cite{suchkov94}) -- the first three bands approximately correspond to the ROSAT bands R1$=0.11-0.284$, R6$=0.73-1.56$, R7$=1.05-2.04$ keV. The rows from top to bottom are for the elapsed times 0.25, 0.5 and 1 Myr. Characteristic features of the X-ray surface brightness distribution can be understood in terms of the density and temperature maps shown in Fig. \ref{fig-multisn-maps}: the brightness mostly follows the product of $n^2T^\beta$, where $\beta$ varies from $\beta\sim 0$ at $T\sim 10^5$ K, to $\beta\simeq -0.5$ at $T\sim 4\times 10^5-10^6$ and to $\beta\simeq 0.5$ at $T\geq 10^7$ K. At $t\simeq 0.25$ Myr, the remnant is already radiative and a dense and cold ($T\sim 10^4$ K) shell has formed. However, there is a thin region just interior to the shell, where the density is less than the shell but the temperature grows up to $T\sim 10^5$ K. This thin region provides an efficient cooling and emission in the lower energy bands. Thus, the upper left map represents emission from this thin layer with a very weak concentration towards the shell. Higher energy bands are less efficiently produced within the bubble, as the internal temperature is either too high ($T\simgt 10^7$ K) or too low ($T\simlt 10^5$ K in the central domain), except the $1.6-8.3$ keV band photons being emitted from the quasi-spherical high-temperature interior (see, Fig. \ref{fig-multisn-maps} for 0.25 Myr). Subsequent evolution clearly shows diminished X-ray emission not only in the higher energy bands, but in the soft band as well, which is the result of evacuation of gas from the interior towards the shell and its cooling.

%%%%%%%%%%%%%%%%%%%%%%%%%%%%%%%%%%%%%%%%%%%%%%%%%%%%%%%%%%%%%%%%%%%%%
\begin{figure*}
\center
\includegraphics[width=16cm]{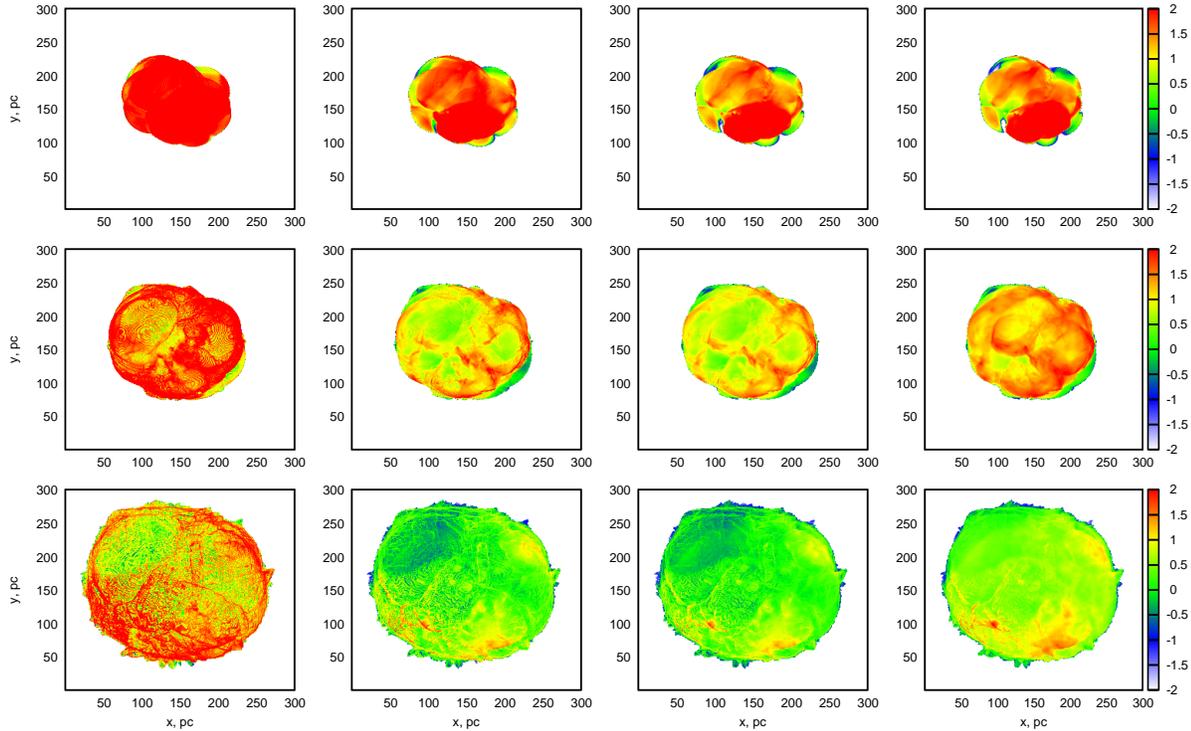}
\caption{
Same as in Fig. \ref{exray_30_evol} for SN rate $4\times 10^{-11}$pc$^{-3}$ yr$^{-1}$ clustered within 90 pc radius.
}
\label{exray-90_evol}
\end{figure*} 
%%%%%%%%%%%%%%%%%%%%%%%%%%%%%%%%%%%%%%%%%%%%%%%%%%%%%%%%%%%%%%%%%%%%%

Explosions with a lower SN rate surprisingly produce more photons in higher  energy bands as seen from Fig. \ref{exray-90_evol}. This is basically connected with the fact that less frequent shock waves inside the bubble evacuate gas outward less efficiently, and leave a larger amount of compressed and relatively hot gas in the form of clumps. Eventually, even at $t=1$ Myr some fraction  of the bubble volume emits at high energies $E>1.8$ keV.

%%%%%%%%%%%%%%%%%%%%%%%%%%%%%%%%%%%%%%%%%%%%%%%%%%%%%%%%%%%%%%%%%%%%%
\begin{figure}
\center
\includegraphics[width=13cm,angle=270]{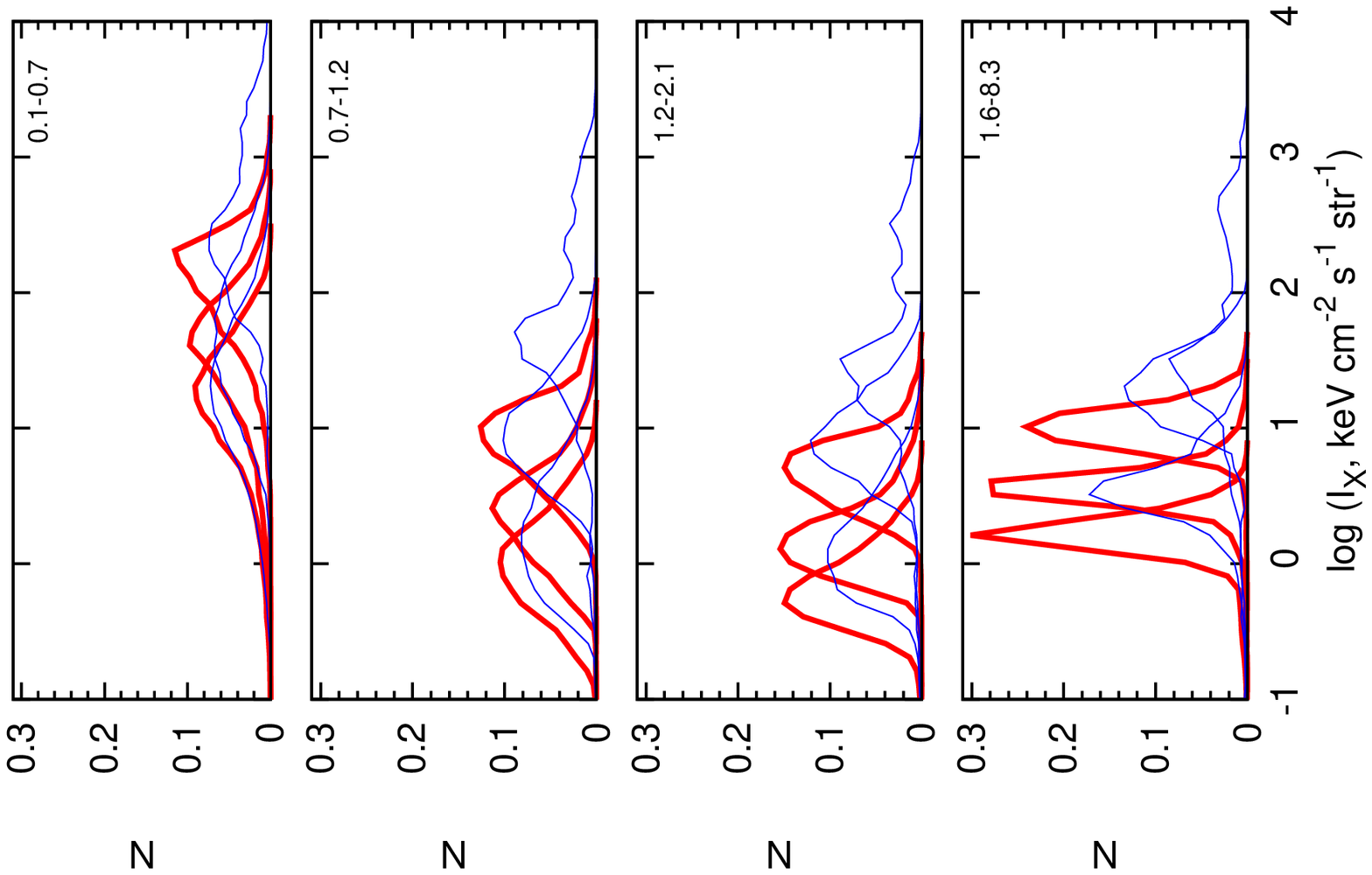}
\caption{
Distribution functions of the surface brightness in X-ray emission for the models shown in Figs.~\ref{exray_30_evol} and \ref{exray-90_evol} for the lower $4\times 10^{-11}$pc$^{-3}$ yr$^{-1}$ (blue thin lines) and the higher $10^{-9}$pc$^{-3}$ yr$^{-1}$ (red thick lines) SN rates clustered within 90 pc and 30 pc, respectively, {for 0.25, 0.5 and 1 Myr (right  to left).}  Panels from the uppermost to the lowermost correspond to X-ray energy bands: $0.1 - 0.3$, $0.7 - 1.2$, $1.2-2.1$ and $1.6-8.3$ keV as shown by legends. 
}
\label{exray-90_band}
\end{figure} 
%%%%%%%%%%%%%%%%%%%%%%%%%%%%%%%%%%%%%%%%%%%%%%%%%%%%%%%%%%%%%%%%%%%%%
{A generic and most obvious feature of the X-ray brightness distribution functions seen in Figure ~\ref{exray-90_band} is that the superbubbles produced by more frequent SNe explosions have narrower distributions in all energy bands and the gas cools faster, while those from slower SNe explosions are wider in photon energy spread and cool slower. This conclusion looks consistent with a narrow density and temperature distributions of a highly compressed gas swept up  mostly into a thin shell by frequent supernovae explosions, and with a more spread and less concentrated gas distribution (generally with lower densities and temperatures) shocked by less frequent supernovae explosions. In addition, in the first case (higher SN rate) the brightness distributions in the high energy bands are narrower than those at lower energies. Moreover, they appear to be squeezed in time. On the contrary, the distributions in the second case (with a lower SN rate) are nearly equally wide. Such a behaviour is consistent with the overall dynamics discussed above. 
}

%----------------------- Section 6 -------------------------------
\section{Conclusions}
We have studied how isolated supernova remnants merge into a collective superbubble depending on the SN rate ambient gas density. 
We have found the following:
\begin{itemize}
 \item The number of SNe exploded in one cooling time ($N_{\rm sn,c}$) is an important factor governing the superbubble growth: the lower is the number, the larger is the fraction of energy lost radiatively and the smaller is the asymptotic radius of the superbubble. 

 \item Superbubbles from clustered SNe explosions pass through a merging phase when individual SN remnants interact and gradually combine into a collective superbubble. For a high SN rate $N_{\rm sn,c}>1$ and a low density environment ($n=1-10$~cm$^{-3}$) the merging phase lasts for a short period ($\simgt 10^4$ yr), {and turns} to an intermediate expansion law with  the {\it effective} radius of the shell $r\sim t^{0.45}$. After about 1~Myr the expansion turns to a steady quasi-wind regime with  $r\sim t^{0.56}$.  
 
 \item In OB-associations with a lower SN rate ($N_{\rm sn,c}<1$) individual SN remnants merge after formation of dense radiative shells. In these conditions a growing superbubble is filled with dense and cool fragments of shells broken in the process of merging. Most of energy is radiated during the merging. Eventually the {\it effective} radius of the superbubble sets on to a wind-like asymptotic  $r(t)\sim t^{0.56}$ on a time scale about 1~Myr.
 
 \item After passing through the intermediate phase, the superbubble eventually settles on to a new power-law wind asymptote with the radius being  smaller than that estimated by using the wind driven bubble law. It results in a significant (factor of a few to one or two orders of magnitude, depending on the ambient density) underestimation of the mechanical luminosity needed to sustain the bubble.
 
 \item Superbubbles with frequent SNe are dominated by the dynamical effects of sweeping the gas outwards and thereby enhancing its cooling in dense external supershells. Bubbles produced by clusters with a lower SN rate, shine brighter in X-ray predominantly, and show a wider brightness distribution function.

\end{itemize}

%----------------------- Section K -------------------------------
\section{Acknowledgements}

\noindent

{The numerical simulations conducted are supported by} the Russian Scientific Foundation (grant 14-50-00043). 
{EV is thankful for support from Ministry of Education and Science (grant 3.858.2017).}
YS is partially supported by the Russian Foundation for Basic Research (grants 15-02-08293, 15-52-45114). 
YS acknowledges hospitality of Raman Research Institute where this work has been finalized.

%----------------------- Section L -------------------------------

%----------------------- Section A -------------------------------

\appendix

\section{A single SN remnant in a {low-density} environment} 

%%%%%%%%%%%%%%%%%%%%%%%%%%%%%%%%%%%%%%%%%%%%%%%%%%%%%%%%%%
\begin{figure*}
\center
\includegraphics[width=8cm]{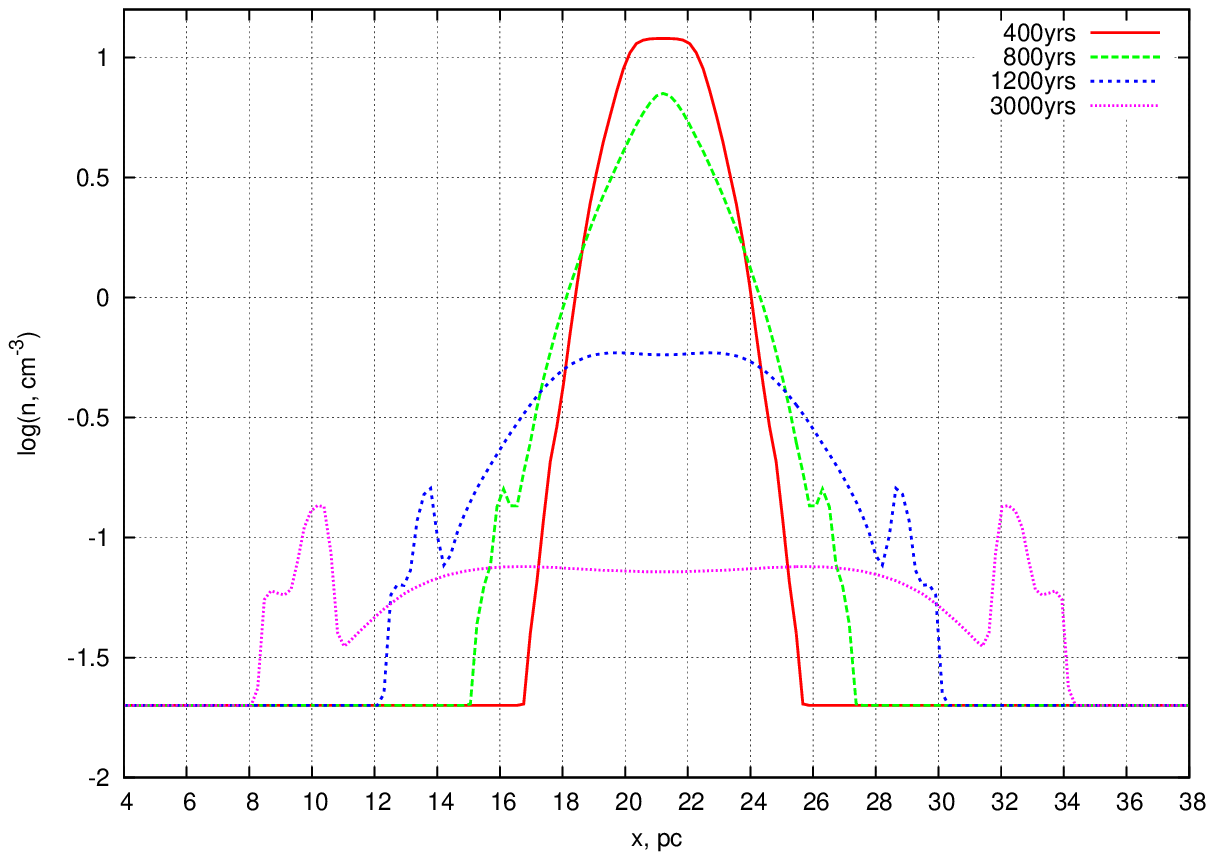}
\includegraphics[width=8cm]{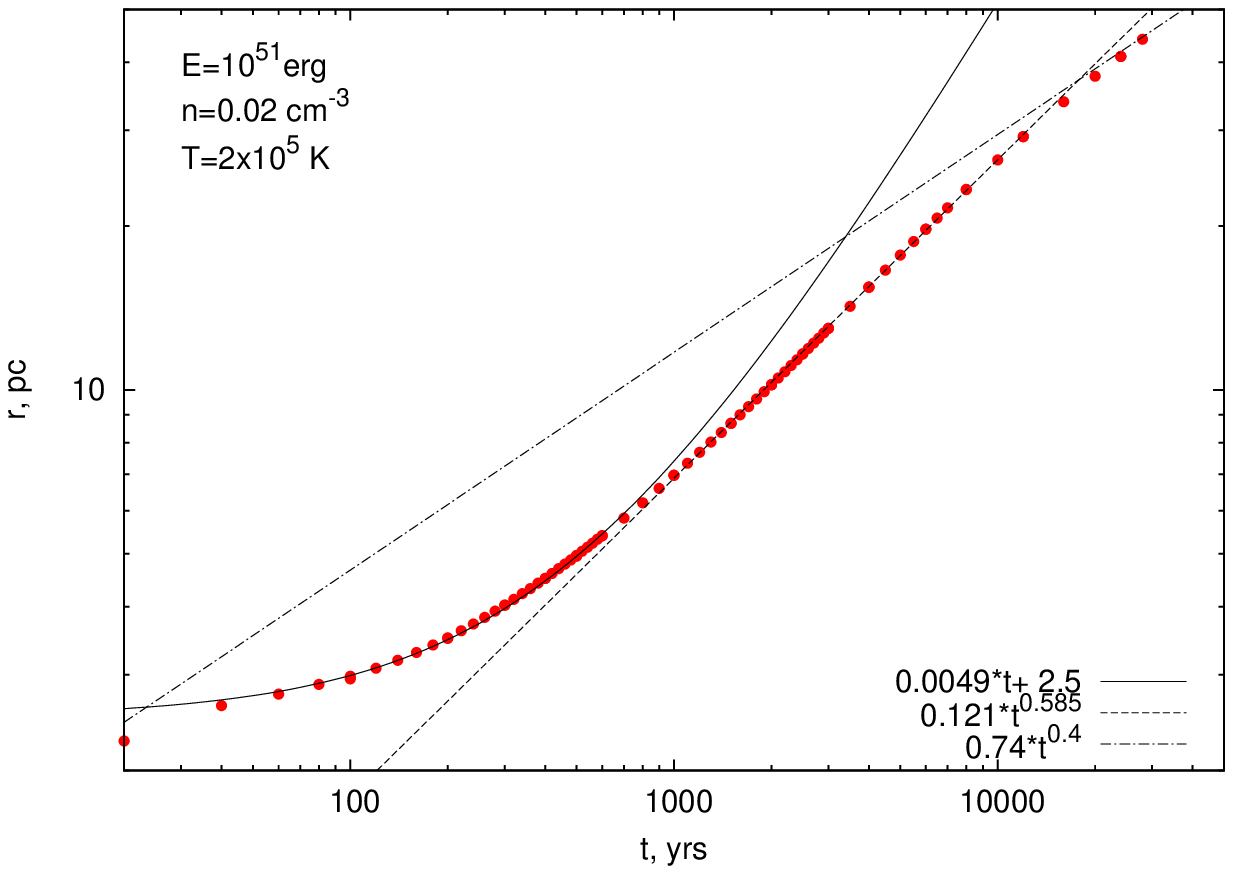}
\break
\includegraphics[width=8cm]{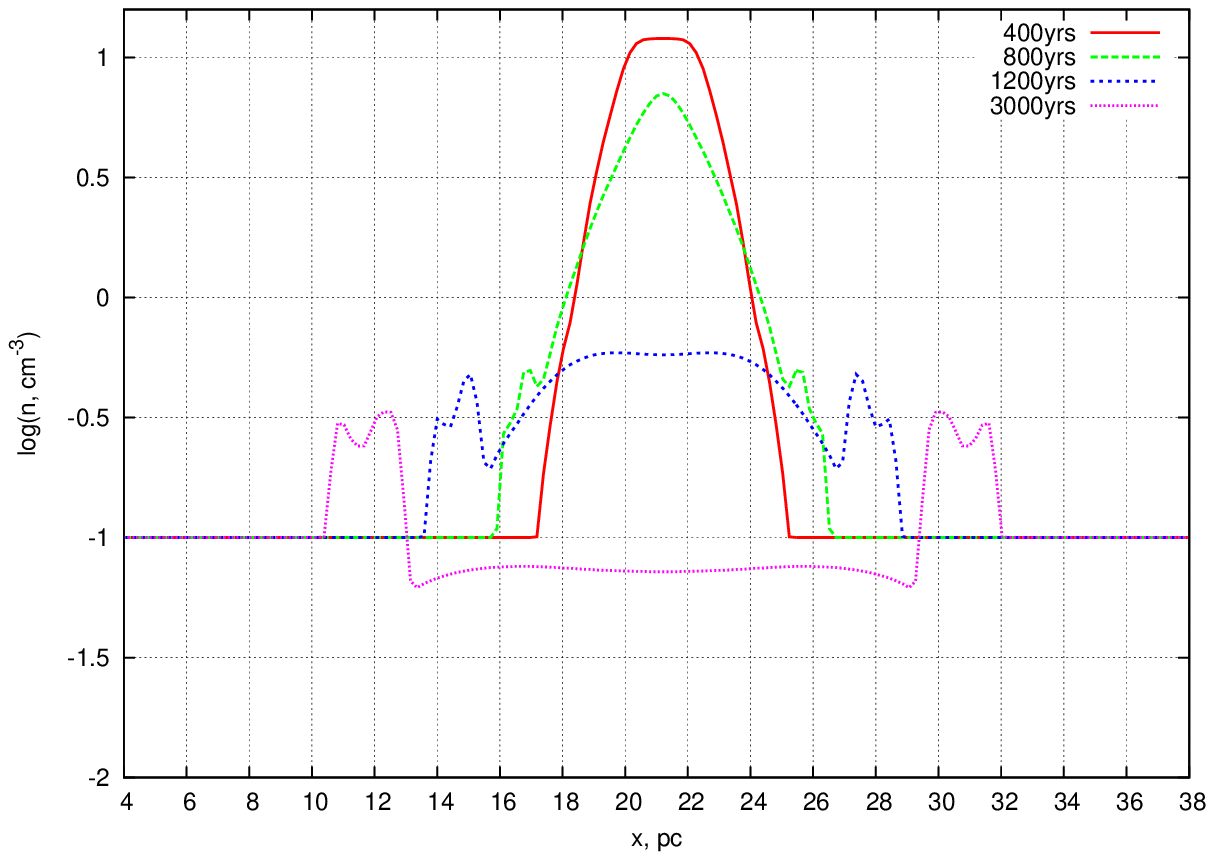}
\includegraphics[width=8cm]{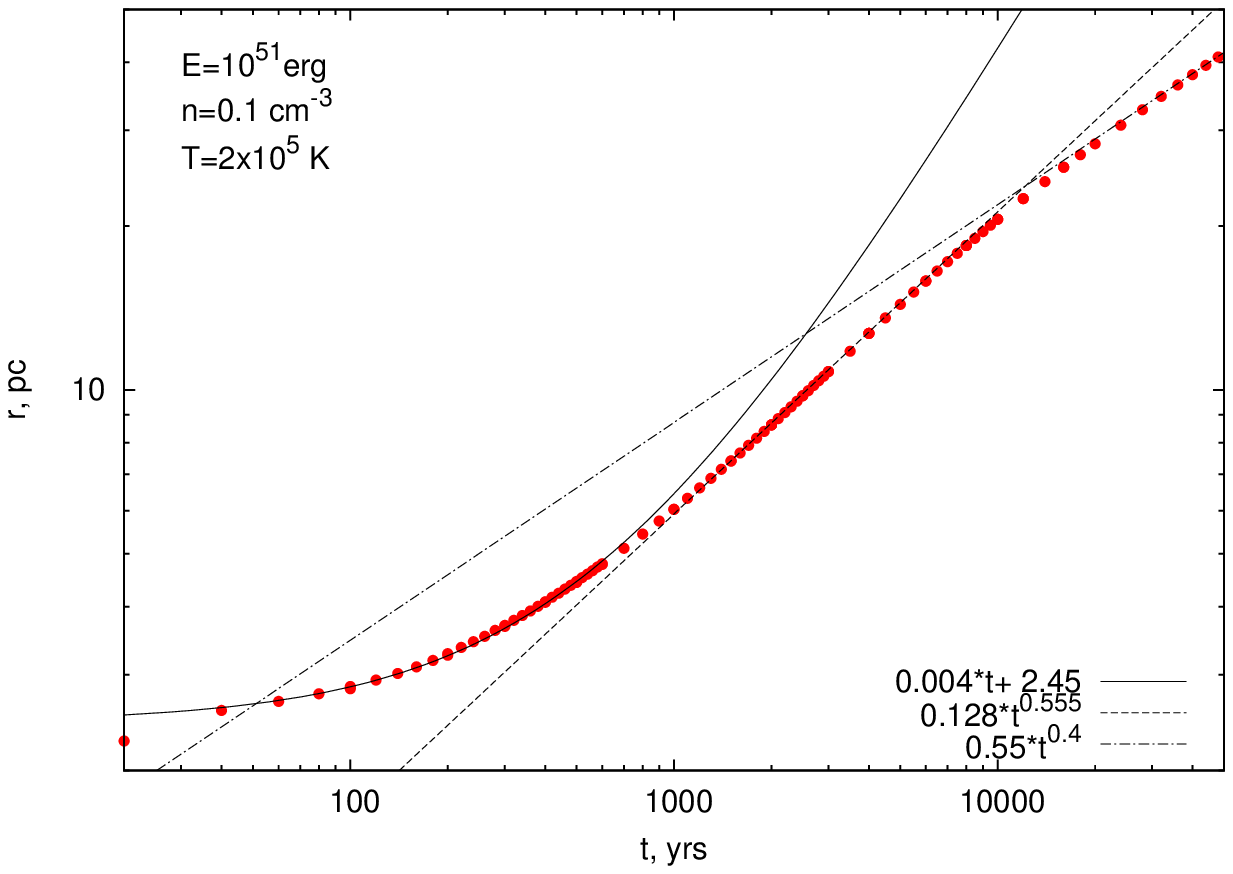}
\break
\includegraphics[width=8cm]{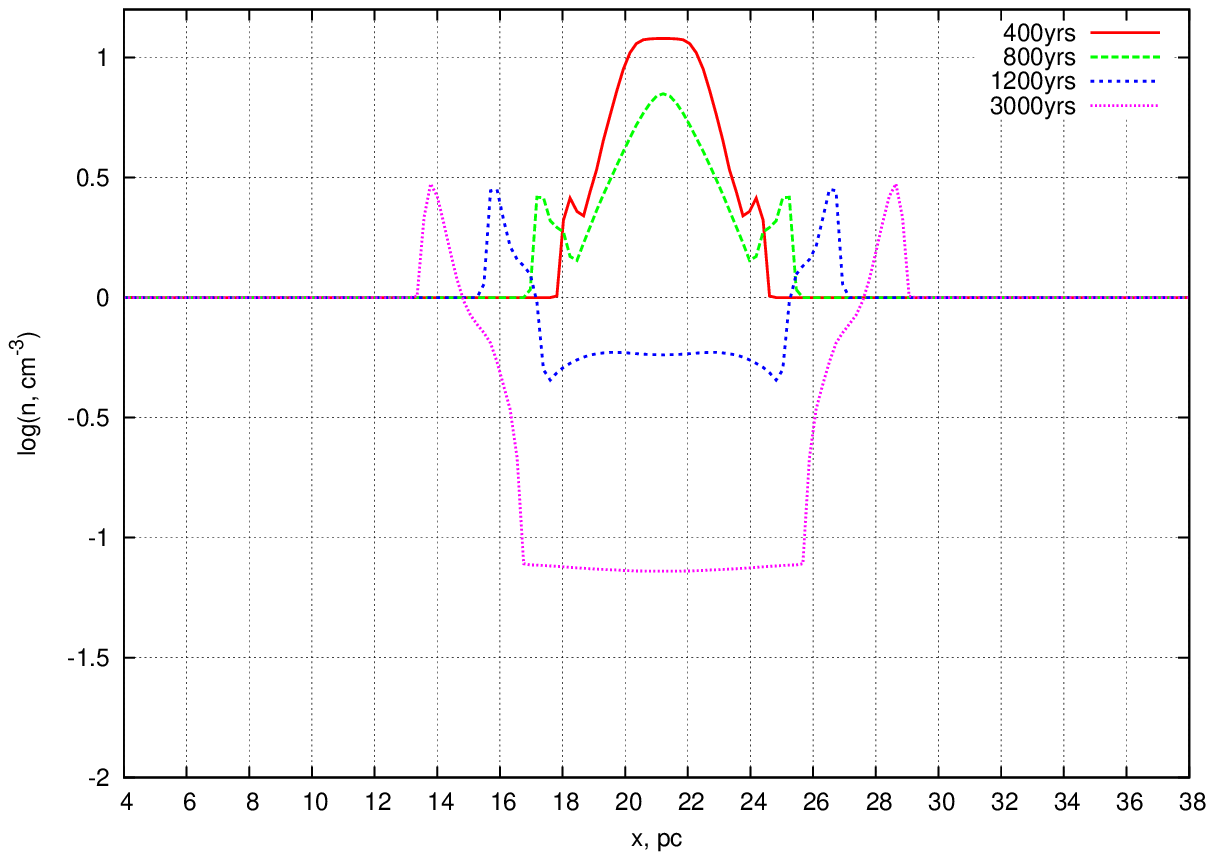}
\includegraphics[width=8cm]{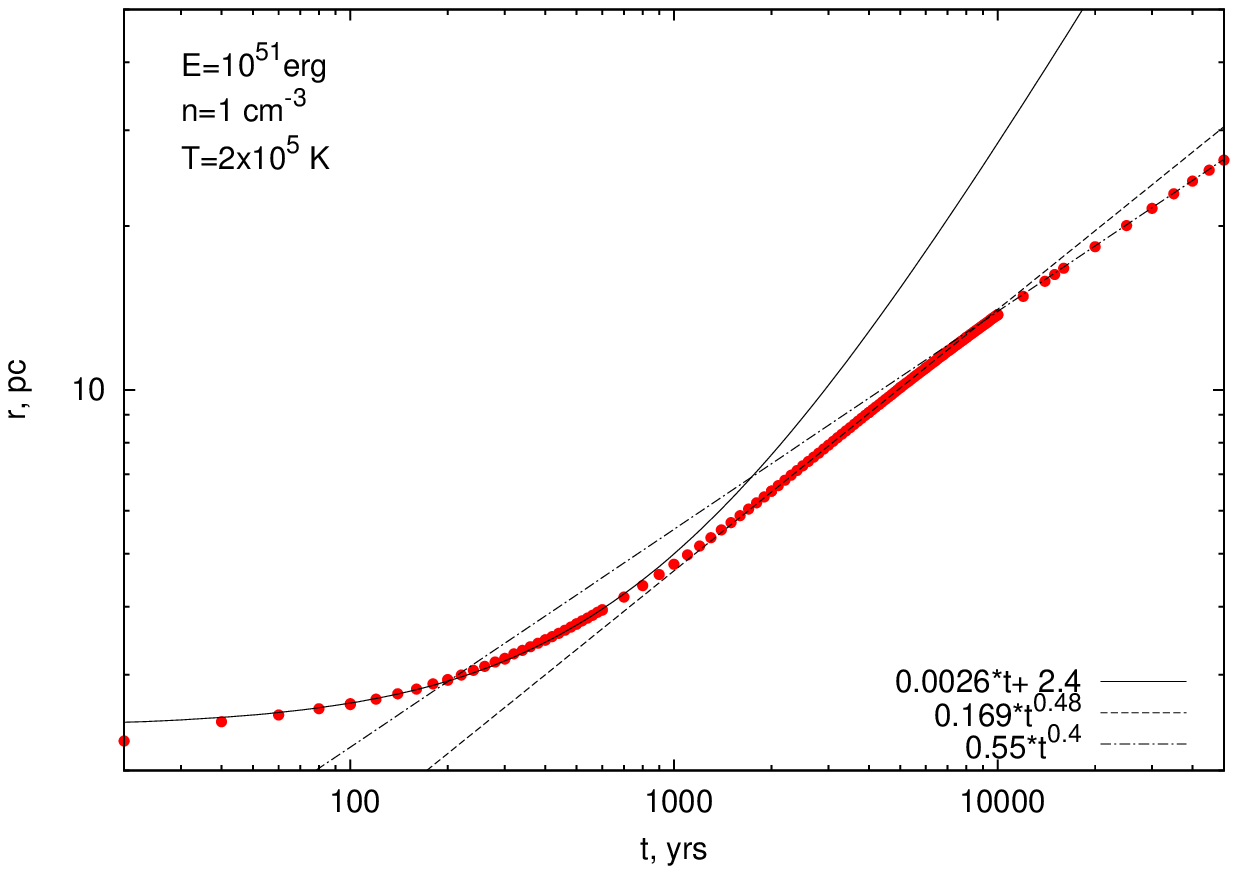}
\caption{
The density profiles (left column) and the evolution of SN shell radius (right column). The number density of the medium is 0.02, 0.1 and 1~cm$^{-3}$ (from top to bottom). 
}
\label{fig-radevol1sn}
\end{figure*}
%%%%%%%%%%%%%%%%%%%%%%%%%%%%%%%%%%%%%%%%%%%%%%%%%%%%%%%%%%%

%%%%%%%%%%%%%%%%%%%%%%%%%%%%%%%%%%%%%%%%%%%%%%%%%%%%%%%%%%
\begin{figure*}
\center
\includegraphics[width=14cm]{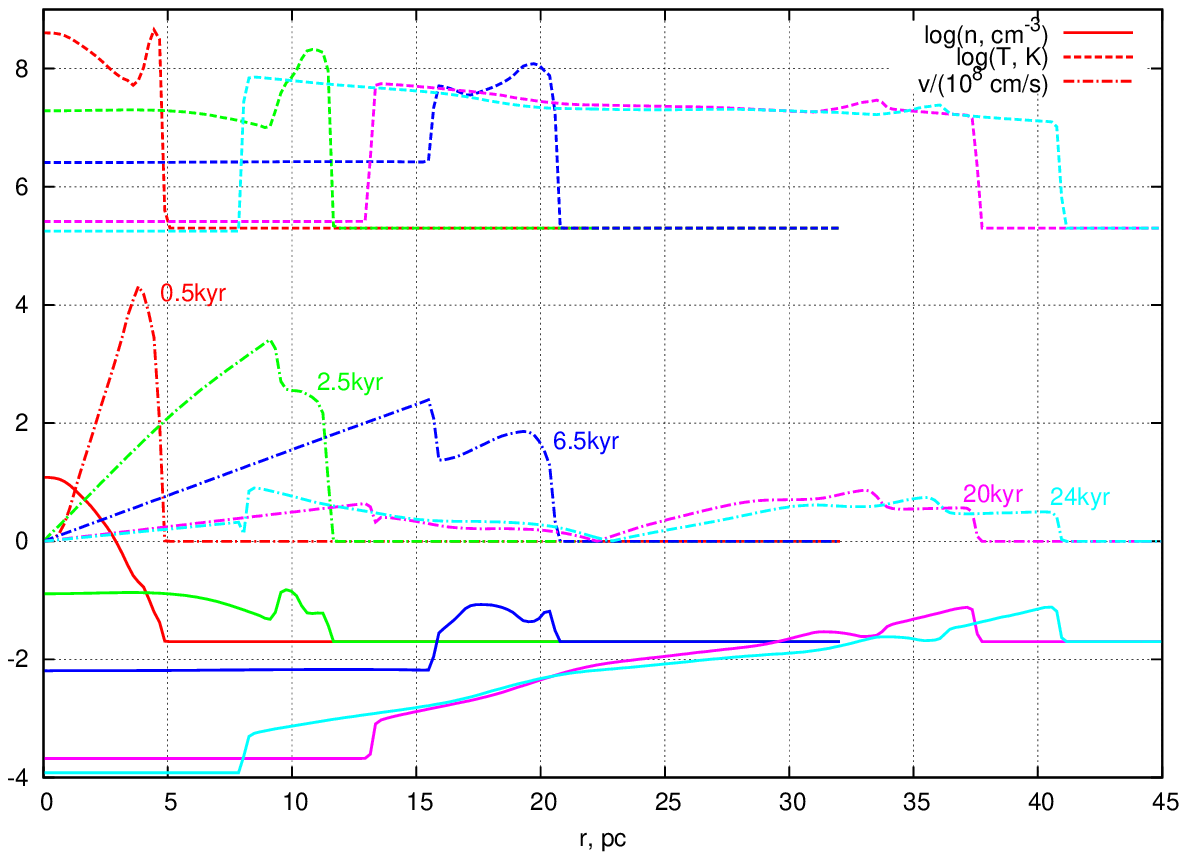}
\caption{
The temperature (dash lines), velocity (dash-dot lines) and density (solid lines) radial profiles at several time moments (see labels) for SN explosion in the medium with number density 0.02~cm$^{-3}$ (see upper panels in Figure~\ref{fig-radevol1sn}). 
}
\label{fig-radpro1sn}
\end{figure*}
%%%%%%%%%%%%%%%%%%%%%%%%%%%%%%%%%%%%%%%%%%%%%%%%%%%%%%%%%%%

In our simulations energy and mass of a SN are injected into a small {volume and the gas density inside turns to be several orders of magnitude higher than the ambient density. Before the shell forms, the dense hot ball expands freely into a tenuous medium. This process can be described as gas expansion into vacuum\footnote{Strictly speaking, fluid dynamics approach violates in the interface separating vacuum and the bulk of gas. However, when gas dynamics far from the interfacial region is concerned the numerical recipes for solving the Riemann problem in presence of vacuum can be found \citep{toro99}.}. The ball radius evolves as $r = r_0 + {2\over \gamma -1 } c_s t$, where $r_0$ is the initial radius, $c_s$ is sound velocity. We assume flat initial radial profiles of density and pressure in the injection region. In general, there is no self-similar solution for this problem \citep[e.g.,][]{zeld-raizer}, besides only some particular initial radial profiles \citep{stanyukovich} which are not applicable in our case.
}

{Linear growth of radius is a part of ejecta-dominated phase \citep[e.g.,][]{gull,mckee99}. The shock gradually collects ambient gas, decelerates and when the mass of swept out gas becomes comparable to the ejecta the expansion {makes a transition} to Sedov-Taylor phase \citep{mckee99}. The duration of the free-expansion phase depends on the ambient density.} 

{Multiple SN explosions efficiently sweep ambient gas out of the central part of the cluster. All subsequent SNe thus expand into a very dilute and hot medium. We consider a single SN explosion into a hot ($T=2\times 10^5$~K) low-density ($n_{\rm amb}=$~0.02, 0.1 and 1~cm$^{-3}$) medium. In order to understand the details of the initial stages, the resolution is taken to be $0.15$~pc, almost $7$ times better than for the majority of runs on later stages. The injected mass and energy of the SN are 10$\msun$ and $10^{51}$~erg.}

{Left panels of Figure~\ref{fig-radevol1sn} present the density profiles at several time moments. It is clearly seen that the shell in front of the {expanding} gas forms practically immediately -- on times $\simeq 600-800$ yr, which is a factor of 3 shorter than that follows from the formal estimate $t_{\rm free}=(3M_{ej}/4\pi\rho)^{1/3}v^{-1}$, $M_{eje}$ is the ejecta mass, $v$ is the initial velocity $v\simeq 3c_s$. It is seen also from the right panel on Fig~\ref{fig-radevol1sn}, where the bubble radius $r(t)$ versus time is shown. Before $t<600$ yr the ejecta expands freely as $r=0.0049t+2.5$ with the constant term approximately equal to the initial ejecta radius and the expansion velocity $v=0.0049$ pc yr$^{-1}$. { This value is only slightly higher than $v_f=2(\gamma -1)^{-1} c_s \simeq 0.0034$ pc yr$^{-1}$ -- the expansion velocity into vacuum, calculated for $T_0=3\times 10^8$~K, which corresponds to the mass $M_{ej}=\msun$ and thermal energy $E=10^{51}$~erg injected into ambient gas with density {$n_{\rm amb}=0.02$}~cm$^{-3}$ within $r_0 = 2$~pc.} After $t\sim (2-3)t_f$, the expansion turns to a quasi-diffusive law $R(t)\propto t^{0.45}$, apparently caused by a series of reversed shock waves; the contribution from numerical diffusion at these time scales is negligible: $(\langle\Delta x\rangle ^2)^{1/2}\sim 2Dt\sim 0.1$ pc at $t\sim 1000$ yr with $D$ determined by the product of the spatial resolution 1 pc and typical sound velocity 40 km s$^{-1}$. Afterwards, the expansion decelerates due to the rarefaction wave propagating inward from the ambient  dilute gas. Eventually after $\simeq 7t\sim t_{\rm free}$ it turns to the Sedov-Taylor solution $r\propto^{2/5}$ (dash-dotted lines). In our models the very initial free-expansion velocity $v_f$ weakly depends on the ambient density as $v_f\propto (1+0.07n_{\rm amb}/M_{ej})^{-1/2}$, here $M_{ej}$ is in units of 10$M_\odot$. Details of the dynamics are seen clearly on Figure~\ref{fig-radpro1sn} where the radial profiles of temperature, velocity and density a SN expansion in the environments with $n=0.02$~cm$^{-3}$ are given. }  

{The structure (density, velocity and temperature profiles) of the bubbles presented here are different from the analytical results of \citet[][ hereafter CC85]{cc85}. However, it has been previously shown the CC85 results are valid for a comparatively large rate of energy deposition. \citet{roy14} showed that CC85 profiles are valid when $\nu_{SN} \ge 3.5 \times 10^{-6}$ pc$^{-3}$ yr$^{-1}$. The SN rate densities considered here are much smaller than this and therefore one does not expect the structure to follow the predictions of CC85, as borne out by our simulations.}

\section{{Pressure-driven phase}} 

{Pressure-driven phase begins at stages when radiation losses come into play, dense shell forms around the hot remnant cavity and the rate of change of momentum is governed by the cavity pressure \citep{mckee77,blinn82}} 

\be 
{d(M\dot R)\over dt}= 4\pi R^2P,
\ee 
{where $M$ is the shell mass, $R$, its radius, $P$, is pressure in the cavity. It can be readily shown that the characteristic time of momentum growth $t_P\sim |M\dot R|/4\pi R^2P\simeq 20 \Delta t$, where $\Delta t$ is time elapsed since the expansion escaped ST phase at $t=t_c$. The shell is being driven by pressure until $t_P$ becomes longer than dynamical time $t_P>t\sim t_c$, meaning $t_P\simeq 0.05 t_c$.  
}

\section{A power law wind-like expansion for a constant SNe rate} 

It can be readily shown that when the SN rate is constant the resulting bubble settles asymptotically on to standard wind regime $R\simeq (Lt^3/\rho)^{1/5}$ with $L\simeq 0.3\nu_{\rm sn}E$ being slightly reduced compare to a commonly assumed $L=\nu_{\rm sn}E$ mechanical luminosity from cumulative SNe explosion.  Indeed, assuming the energy injected continuously into the bubble to be written as 
\be 
L=L_0-{3\over 2}k_{\rm B}Tn\dot V,
\ee 
where $L_0=\nu_{\rm sn}E$, $V$ is the volume occupied by the shell, we explicitly assume here that thermal energy of shocked gas $3k_{\rm B}T/2=9m_{\rm H}\dot R^2/32$ is completely lost radiatively when it passes through the shell. Accounting that the postshock flow inside the bubble is subsonic we can seek the solution for the shell radius
\be 
{4\over 3}\pi\rho R^3\ddot R=2{\int Ldt\over R}-4\pi\rho\dot R^2R^2,
\ee 
as a power-law $R(t)\propto t^\alpha$ to find $R\simeq (0.3L_0t^3/\rho)^{1/5}$.

\end{document}